\documentclass[journal,twocolumn,final]{IEEEtran}
\usepackage[final]{graphicx}
\usepackage[dvips,all,color]{xy}
\xyoption{arc}
\usepackage{subfigure}
\usepackage{amsmath,amssymb}
\usepackage[svgnames,x11names]{xcolor}
\usepackage{dsfont}
\usepackage[linesnumbered,ruled,vlined,longend]{algorithm2e}
\usepackage{mathrsfs}
\usepackage{times}
\usepackage{stmaryrd}
\usepackage{ifsym}
\usepackage{flushend}
\usepackage{here}
\usepackage{fancyhdr}
\usepackage{psfrag}
\usepackage{shading}
\usepackage{ifthen}
\usepackage[normalem]{ulem}
\newcommand{\forloop}[5][1] { \setcounter{#2}{#3} \ifthenelse{#4} { #5 \addtocounter{#2}{#1} \forloop[#1]{#2}{\value{#2}}{#4}{#5} }  Else { } }
\DeclareMathAlphabet{\mathpzc}{OT1}{pzc}{m}{it}

\newtheorem{lem}{Lemma}
\newtheorem{prop}{Proposition}
\newtheorem{defn}{Definition}

\newtheorem{rem}{Remark}
\newtheorem{notn}{Notation}

\newcommand{\Gnet}{\mathds{G}_{\textrm{\sffamily\textsc{N}}}}
\newcommand{\cur}{\textrm{}}
\newcommand{\Sn}{\textrm{\sffamily\textsc{Sink}}}

\newcommand{\Crd}{\textrm{\sffamily\textsc{Card}}}

\newcommand{\eg}{\geqq_{\textbf{\texttt{(Elementwise)}}}}
\newcommand{\Q}{\mathscr{P}}
\newcommand{\Pitilde}{\widetilde{\Pi}}
\newcommand{\mudble}[2]{\widehat{\nu}^\infty_{\left(#1,#2\right)}}
\newcommand{\nuinf}[2][\theta]{\widehat{\nu}^\infty_{#1} #2}

\newcommand{\ones}{\mathbf{e}}
\newcommand{\myar}{\ar@[|(2.5)]}
\newcommand{\myarT}{\ar@[|(3.5)]}
\newcommand{\myarL}{\ar@[|(1.5)]}
\newcommand{\upd}{\mathds{U}}
\newcommand{\Tc}{\mathds{T}_c}
\newcommand{\cgather}[2][0pt]{\begingroup\setlength\abovedisplayskip{#1}\setlength\belowdisplayskip{#1}\begin{gather} #2 \end{gather}\endgroup}
\newcommand{\calign}[2][0pt]{\begingroup\setlength\abovedisplayskip{#1}\setlength\belowdisplayskip{#1}\begin{align} #2 \end{align}\endgroup}
\newcommand{\Mblue}{\color{MidnightBlue}}
\newcommand{\Nblue}{\color{DodgerBlue}}
\newcommand{\Dgreen}{\color{DarkGreen}}
\newcommand{\BRed}{\color{DarkRed}}

\newcommand{\red}{\color{DarkRed}}
\newcommand{\black}{\color{black}}

\title{ GODDeS: \uline{G}lobally $\epsilon$-\uline{O}ptimal Routing Via \uline{D}istributed \uline{De}cision-theoretic  \uline{S}elf-organization  
}
\author{ 
Ishanu~Chattopadhyay~\IEEEauthorrefmark{1}%, Yicheng~Wen~\IEEEauthorrefmark{2} and Asok~Ray~\IEEEauthorrefmark{3}
\IEEEcompsocitemizethanks{
\IEEEcompsocthanksitem\IEEEauthorrefmark{1} Corresponding Author, email: ixc128@psu.edu
\IEEEcompsocthanksitem Mechanical Engineering, The Pennsylvania State University,  USA
\IEEEcompsocthanksitem Supported in part by the U.S.  Army Research Office (W911NF-07-1-0376) and  the Office of Naval Research (N00014-09-1-0688). 
} 
} 
\begin{document}
\maketitle
%##############################################################
\allowdisplaybreaks{
%##############################################################
\begin{abstract}
This paper introduces GODDeS: a fully distributed self-organizing decision-theoretic routing algorithm  designed to effectively exploit high quality paths in  lossy  ad-hoc wireless environments, typically with a large number of nodes.
The routing problem is modeled as an optimal control problem for a decentralized Markov Decision Process, with links  characterized by locally known packet drop probabilities that either remain constant on average or change slowly. The equivalence of this optimization problem  to that of performance maximization of an explicitly constructed  probabilistic automata allows us to effectively apply the theory of quantitative measures of probabilistic regular languages, and design a distributed highly efficient solution approach that attempts to minimize source-to-sink drop probabilities across the network.
 Theoretical results provide rigorous guarantees on global performance, showing that the algorithm achieves near-global optimality, in polynomial time.
It is also argued that GODDeS is significantly congestion-aware, and exploits multi-path routes optimally. Theoretical development is supported by high-fidelity network simulations.
\end{abstract}
%##############################################################
%##############################################################
% 
\section{Introduction \& Motivation}
The routing problem has been widely studied in the context of ad-hoc wireless networks, and reported algorithms can be broadly classified as follows. A routing protocol is pro-active (DBF ($e.g.$ Distributed Bellman-Ford)~\cite{Bertsekas1987a} and DSDV (Highly Dynamic Destination-Sequenced Distance Vector routing)~\cite{Perkins2001b}), if fresh destination lists and their routes are maintained by periodically distributing routing tables; it is reactive ($e.g.$ AODV (Ad-hoc On-demand Distance Vector)~\cite{Perkins1999} and DSR (Dynamic Source Routing)~\cite{Johnson2001}) if routes are computed if and when necessary by flooding the network with Route Request packets. Pro-active protocols suffer from expensive route maintenance and slow reaction to topology changes, while reactive methods  have high latency in  discovery and  induce  congestion due to periodic flooding.
Hybrid protocols attempt to combine advantages of both  philosophies $e.g.$ HRPLS (Hybrid Routing Protocol for Large Scale Mobile Ad Hoc Networks with Mobile Backbones)~\cite{PAKG06} and HSLS (Hazy Sighted Link State routing protocol)~\cite{KDP05}. Protocols may also be classified as being either distance-vector or link-state driven. In the former case, the  computed distance to all nodes  is  is exchanged with neighbors ($e.g.$ DSDV, AODV); while in the latter computed distances to the neighbors  is exchanged with all nodes ($e.g.$ OLSR (Optimized Link State Routing)~\cite{Jacquet2001}, ZHLS (Zone-Based Hierarchical Link State)~\cite{JoaNg1999a}). 
Link state protocols maintain better  Quality Of Service (QOS), but  suffer from poor scalability. Distance vector protocols have less control traffic, but maintaining QOS is more difficult.
 Other approaches use 
geographic,  or power  information, and in the context of sensor networks, query based  routing strategies ($e.g.$ Directed Diffusion~\cite{IGE00}) have been proposed. 
% 
% 
% Routing protocols may be designed to merely find feasible routes that 
% put emphasis on rapid  discovery or robust  maintenance, or may be optimized to achieve pertinent objectives such as minimizing hop counts or communication overheads or maximizing network life ($e.g.$ energy-aware routing).
%  See [ref]  for a   survey  of reported protocols.
% 

Reported ad hoc routing protocols for wireless networks primarily focus on 
% strategies to cope
% Much of the recent work in ad hoc routing protocols for wireless networks [24,14,25] has focused on 
% coping
%  with 
node mobility, rapidly changing topologies, overhead, and scalability; with little attention  paid to finding high-quality paths in the face of
lossy wireless links. An implicit assumption is that links either work well
or don’t work at all; which is not  reasonable  in the wireless case where many
 links have intermediate loss ratios. This problem has  been partially addressed by designing new 
quality-aware metrics such as the expected transmission count (ETX)~\cite{CABM05}, where the authors correctly note ``{\itshape minimizing  hop-count maximizes the distance traveled
by each hop, which is likely to minimize signal strength and
maximize the loss ratio}''. Even if the best route is  one with minimal
hop-count, there may be many routes (particularly in dense networks)
of the same minimum length with widely varying qualities; arbitrary choice made by most minimum hop-count metrics
is not likely to select the best. The problem is also crucial in  multi-rate
networks~\cite{AHR04}, where the routing protocol must select from the set of available links. While in single-rate networks
all links are equivalent, in multi-rate networks each available
link may operate at a different rate. Thus the routing protocol
is presented with a  complex trade-off decision: {\itshape Long distance links take fewer hops, but  the links  operate slower; short links can operate at
high rates, but more hops are required.} 

In this paper, we give a theoretical solution to this potentially large-scale decision problem via formulating a probabilistic routing policy that very nearly minimizes 
the end-to-end packet drop probabilities.
In particular,  the  routing problem is   modeled and solved  as an  optimal control problem for  a Decentralized Markov Decision Process (D-MDP). 
Extensively used for centralized decision making in
stochastic environments, Markov decision processes (MDPs) have been  recently,
extended  to decentralized multi-agent settings~\cite{Bn00}. In the context of ad-hoc routing, we begin by assuming  that the communication links are  imperfect, and are being characterized by  locally known drop probabilities.  The mean or expected values of the link-specific drop probabilities, and the network topology is assumed to be are either constant or changing over  a time scale which is significantly slower compared to that of the communication dynamics. We then seek  local routing decisions that maximize throughput in the sense of minimizing the source-to-sink probability of packet-drops. The Markov structure emerges, since we assume that the local link-specific drop probabilities  are independent of the history of sequential link traversal by individual packets.

The results developed in this paper effectively resolve the issues described above (and does more, actually attaining near global optimality); and would seem to be a straightforward solution scheme. Nevertheless, to the best of the author's knowledge, such an approach has not been previously investigated. The reason for this apparent neglect (which also highlights the key theoretical contribution of this paper) is as follows:
Recent investigations~\cite{Bn00,BGIZ02} into the solution complexity of  decentralized Markov decision processes  have shown that the problem is exceptionally hard  even for two agents;  illustrating a fundamental
divide between centralized and decentralized control of MDP. In contrast
to the centralized approach, the decentralized case provably does not admit
polynomial-time algorithms. Furthermore, assuming $\textrm{EXP} = \textrm{NEXP}$, the problems require super-exponential time to solve in the worst case.
Such negative results do not preclude the possibility of obtaining {\itshape near-optimal} solutions efficiently. This is precisely what we achieve in this paper, in the context of the 
routing problem. We show that a highly efficient, fully distributed, decision algorithm can be designed that effectively solves the distributed MDP such that the control policy, on convergence, is within an $\epsilon$ bound of the global optimal. Furthermore, one can freely choose the error bound $\epsilon$ (and make it as small as one wishes), with the caveat that the convergence time 
increases (with no finite upper bound) with decreasing $\epsilon$.

We call this algorithm GODDeS (\uline{G}lobally $\epsilon$-\uline{O}ptimal Routing Via \uline{D}istributed \uline{De}cision-theoretic  \uline{S}elf-organization).
Instead of using a standard MDP formulation, we use a problem representation based on Probabilistic Finite State Automata (PFSA), which allows us to set up the decision problem as that of performance maximization of PFSA, and obtain solutions using the recently reported  quantitative measures of probabilistic regular languages~\cite{CR07}. This shift of modeling paradigm is the quintessential insight that 
allows one to achieve near-global optimality in polynomial time.  Theoretical results also establish that GODDeS is highly scalable,  optimally take advantage of existing multi-path routes, and is expected to be significantly congestion-aware. For simplicity of exposition, a single sink is considered throughout the paper. This is not a serious restriction, since the results carry over to the general case with ease. The resulting algorithm is both pro-active and reactive, but not  in the usual sense of reported hybrid protocols. It uses both distance-vector (in a generalized sense via the language-measure construction) and link-state information, and uses local multi-cast to forward messages; optimally taking advantage of multi-path routing. 

% \regtotcounter{section}
The rest of the  paper is organized in   six sections. Section~\ref{sec2} briefly summarizes the theory of quantitative measures of probabilistic regular languages, and the pertinent approaches to centralized performance maximization of PFSA. Section~\ref{sec3} develops the PFSA model of an ad-hoc network, and  Section~\ref{sec4} presents the key theoretical development for decentralized PFSA optimization. Section~\ref{sec5} validates the theoretical development with high fidelity simulation results on the NS2 network simulator, and discusses the key properties and characteristics for the proposed routing algorithm. The paper is summarized and concluded in Section~\ref{sec7} with recommendations for future work.
% 
% 
% 
% 
% % \clearpage ~\cite{Bn00,BGIZ02}
% \begin{figure}[t]
% \centering
% \vspace{60pt}
% \psfrag{81}[l][c][1][90]{\scriptsize  \txt{GB}}
% \psfrag{87}[l][c][1][90]{\scriptsize  \txt{DBF}}
% \psfrag{94}[l][l][1][90]{\scriptsize  \txt{DSDV}}
% \psfrag{95}[l][c][1][90]{\scriptsize  \txt{LMR, WRP}}
% \psfrag{96}[l][c][1][90]{\scriptsize  \txt{DSR, ABR, SSA}}
% \psfrag{97}[l][c][1][90]{\scriptsize  \txt{CGSR, ZRP, AODV}}
% \psfrag{98}[l][r][1][90]{\scriptsize  \txt{$\phantom{X}$\\CEDAR, OLSR \\ZHLS, STAR$\phantom{.}$$\phantom{.}$$\phantom{.}$\\ LEACH$\phantom{X}$$\phantom{X}$$\phantom{X}$$\phantom{X}$ }}
% \psfrag{01}[l][r][1][90]{\scriptsize  \txt{$\phantom{.}$\\BRP, IARP, IERP}}
% \psfrag{02}[l][r][1][90]{\scriptsize  \txt{$\phantom{.}$\\PLBR}}
% \psfrag{04}[l][r][1][90]{\scriptsize  \txt{$\phantom{.}$\\Directed Diffusion}}
%  \includegraphics[width=3.5in]{Figures/tline}
% \caption{Timeline of Ad-hoc Protocols (Non-exhaustive)}\label{figtimeline}
% \end{figure}
% 
% {\BRed Reactive/Proactive}, {\BRed Distance-vector/Link-state}, Hierarchical/Clustered/Flat, {\BRed Uniform/Non-uniform}, {\BRed Past/Prediction}, {\BRed Local Multicast}/Flooding, {\BRed Multipath}/Singlepath
% 
\section{Background: Language Measure Theory}\label{sec2}
This section summarizes the concept of signed real measure of probabilistic regular
languages, and its application in performance optimization of 
probabilistic finite state automata (PFSA)~\cite{CR07}.
A string over an alphabet ($i.e.$ a non-empty finite set) $\Sigma$ is a finite-length sequence of symbols from $\Sigma$~\cite{HMU01}. 
The Kleene closure of $\Sigma$, denoted by $\Sigma^*$, is the set of all finite-length strings of symbols including the null string $\epsilon$.
The string $xy$ is the  concatenation of strings $x$ and $y$, and the null string $\epsilon$ is the identity element of the concatenative monoid.
\begin{defn}[PFSA]\label{defPFSA}
A PFSA $G$ over an alphabet $\Sigma$ is a sextuple $(Q,\Sigma,\delta,\widetilde{\Pi},\chi,\mathscr{C})$, where $Q$ is a set of states, $\delta:Q\times\Sigma^\star \rightarrow Q$ is
the (possibly partial) transition map; $\widetilde{\Pi}: Q\times \Sigma \rightarrow [0,1]$ is an output mapping, known as the probability morph function that
specifies the state-specific symbol generation probabilities and satisfies
$\forall q_i \in Q, \sigma \in \Sigma, \widetilde{\Pi}(q_i,\sigma) \geqq 0$, and $ \sum_{\sigma\in\Sigma}\widetilde{\Pi}(q_i,\sigma)=1$, the state characteristic function $\chi:Q \rightarrow [-1, 1]$ 
assigns a signed real weight to each state, and $\mathscr{C}$ is the set of controllable transitions that can be disabled (Definition~\ref{defcontapp}).
\end{defn}
\begin{defn}[Control Philosophy]\label{defcontapp}
If $\delta(q_i,\sigma) = q_k$, then the \textit{disabling}
of  $\sigma$ at 
$q_i$ prevents the state transition from $q_i$ to $q_k$.  Thus, disabling a transition $\sigma$ at a  state $q$ replaces 
the original transition with a  self-loop  with identical occurrence probability, $i.e.$ we now have $\delta(q_i,\sigma) = q_i$. 
Transitions that can be so disabled are 
\textit{controllable}, and belong to the set $\mathscr{C}$.
\end{defn}\vspace{0pt}
\begin{defn}\label{Lgen}
The language $L(q_i)$ generated by a PFSA $G$ initialized at the
state $q_i\in Q$ is defined as:
$    L(q_i) = \{s \in \Sigma^* \ | \ \delta(q_i, s) \in Q \}$
Similarly, for every $q_j\in Q$,  $L(q_i, q_j)$ denotes the set of all
strings that, starting from the state $q_i$, terminate at the
state $q_j$, i.e.,
$L(q_i,q_j) = \{ s \in \Sigma^* \ | \ \delta(q_i, s) = q_j \in Q \}$
\end{defn}
% % % %
\begin{defn}[State Transition Matrix]\label{pifn}The state transition probability matrix $\Pi \in [0,1]^{\Crd(Q) \times \Crd(Q)}$,
for a given PFSA is defined as:
$\forall q_i, q_j \in Q, \Pi_{ij} =
     \sum_{\sigma\in\Sigma \ \mathrm{s.t.} \  \delta(q_i,\sigma)=q_j } \widetilde{\Pi}(\sigma, q_i)$
Note that $\Pi$ is a square non-negative stochastic matrix~\cite{BR97}, where $\Pi_{ij}$ is the probability of transitioning from  $q_i$ to $q_j$.
\end{defn}\vspace{0pt}
\begin{notn}
We use matrix notations interchangeably for the morph function $\widetilde{\Pi}$. In particular,
$\widetilde{\Pi}_{ij} = \widetilde{\Pi}(q_i,\sigma_j)$ with $ q_i \in Q, \sigma_j \in \Sigma$.
Note that $\widetilde{\Pi} \in [0,1]^{\Crd(Q) \times \Crd(\Sigma)}$ is not necessarily square, but each row sums up to unity.
\end{notn}
A signed real measure~\cite{R88} $\nu^i:{2^{L(q_i)}} \rightarrow
\mathbb{R}\equiv(-\infty,+\infty)$ is constructed on the
$\sigma$-algebra $2^{L(q_i)}$~\cite{CR07}, implying that every singleton string set $\{ s \in L(q_i) \} $ is a measurable set. 
\begin{defn}[Language Measure]\label{measurefn}Let $\omega \in L(q_i, q_j)\subseteq 2^{L(q_i)}$. The signed
real measure $\nu^i_\theta$ of every singleton string set $ \{ \omega
\} $ is defined as:
$\nu^i_\theta(\{\omega \})\triangleq\theta (1-\theta)^{\vert \omega \vert}\widetilde{\Pi}(q_i,\omega)\chi(q_j)$.
For every choice of the parameter $\theta \in (0,1)$, the signed real measure of a sublanguage $L(q_i,q_j) \subseteq
L(q_i)$ is defined as:
%\begin{equation}\label{measureEqn}
$ \nu^i_\theta(L(q_i, q_j)) \triangleq \sum_{\omega\in
L(q_i, q_j)} \theta (1-\theta)^{\vert \omega \vert}\widetilde{\Pi}( q_i,\omega)\chi_j
$. Similarly, the measure of $L(q_i)$, is defined as 
$\nu^i_\theta(L(q_i)) \triangleq \sum_{q_j \in Q}
\nu^i_\theta(L_{i,j})$.
\end{defn}
\begin{notn}
 For a given PFSA, we interpret the set of measures $\nu^i_\theta(L(q_i))$ as a real-valued vector of length $\Crd(Q)$ and  denote $\nu^i_\theta(L(q_i))$ as $\nu_\theta \vert_i$.
\end{notn}
The language measure can be expressed vectorially:
\begin{gather}\label{eqmesd}
 \nu_\theta = \theta \big [ \mathbb{I} - (1-\theta)\Pi \big ]^{-1} \chi 
\end{gather}
The inverse  exists for $\theta \in (0,1]$~\cite{CR07}.
\begin{rem}[Physical Interpretation]
In the limit of $\theta \rightarrow 0^+$, the language measure of singleton strings can be interpreted to be product of the conditional generation probability of the string, and the 
characteristic weight on the terminating state. Hence, smaller the characteristic, or smaller the probability of generating the string, smaller is its measure. Thus, if the
characteristic values are chosen to represent the control specification, with more positive weights given to more desirable states, then the measure represents how \textit{good} the particular string is with respect to the given specification, and the given model. The limiting language measure $\nu_0\vert_i = \lim_{\theta \rightarrow 0^+}\theta \big [ \mathbb{I} - (1-\theta)\Pi \big ]^{-1} \chi \big \vert_i$
sums up the limiting measures of each string starting from $q_i$, and thus captures how \textit{good} $q_i$ is, based on not only its own characteristic, but on how \textit{good} are the  strings  generated in  future from $q_i$. It is thus a quantification of the impact of $q_i$, in a probabilistic sense, on future dynamical evolution~\cite{CR07}.
\end{rem}
\begin{defn}[Supervisor]A supervisor disables a subset of the set $\mathscr{C}$ of
controllable transitions and hence there is a bijection
between the set of all possible supervision policies and the
power set $2^{\mathscr{C}}$. \end{defn}

Language measure  allows a quantitative comparison of
different supervision policies.
\begin{defn}[Optimal Supervision Problem]\label{pdef}
Given a PFSA
$G=(Q,\Sigma,\delta,\widetilde{\Pi},\chi,\mathscr{C})$, compute a supervisor disabling 
$\mathscr{D}^\star \subseteq \mathscr{C}$, s.t. $
\nu^{\star}_0 \eg \nu^{\dag}_0 \ \
\forall \mathscr{D}^{\dag} \subseteq \mathscr{C} $ where
$\nu^{\star}_0$, $\nu^{\dag}_0$ are
the limiting measure vectors of  supervised plants $G^{\star}$, $G^{\dag}$ under $\mathscr{D}^{\star}$, $\mathscr{D}^\dag$ respectively.
\end{defn}
\begin{rem}\label{remtheta}
The solution to the optimal supervision problem  is obtained
in \cite{CR07} by designing an optimal policy using $\nu_\theta$ with $\theta \in (0,1)$. To ensure
that the computed optimal policy coincides with the one for
$\theta \rightarrow 0^+$, the authors choose a \textit{small, but non-zero}
value for $\theta$ in each iteration step of the design
algorithm. 
To address numerical issues, algorithms reported in \cite{CR07}
computes how small a $\theta$ is actually required, $i.e.$,
computes the critical lower bound $\theta_\star$.
Moreover the solution obtained is optimal, unique, efficiently computable, and maximally permissive among policies with maximal performance.
\end{rem}
Language-measure-theoretic optimization is \textit{not a search based approach}. It is an iterative sequence of combinatorial manipulations, that 
monotonically improves the measures, leading to  element-wise maximization of $\nu_\theta$ (See \cite{CR07}).
It is shown in \cite{CR07} that
\begin{gather}
 \lim_{\theta \rightarrow 0^+}  \theta \big [ \mathbb{I} - (1-\theta)\Pi \big ]^{-1} \chi = \Q \chi
\end{gather}
where the $i^{th}$ row of $\Q$ (denoted as $\wp^i$) is the stationary probability vector for the PFSA initialized at state $q_i$. 
In other words, $\Q$ is the Cesaro limit of the stochastic matrix $\Pi$, satisfying
% \begin{gather}
$ \Q = \lim_{k\rightarrow\infty}\sum_{j=0}^k\Pi^k$~\cite{BR97}.
% \end{gather}
% 
\begin{prop}[See \cite{CR07}]\label{prop1}
Since the optimization maximizes the language measure element-wise for $\theta \rightarrow 0^+$, it follows that 
for the optimally supervised plant, the standard inner product $\langle \wp^i,\chi\rangle$ is maximized, irrespective of the starting state $q_i \in Q$.
\end{prop}
\begin{notn}
The optimal $\theta$-dependent measure for a PFSA is denoted as $\nu^\star_\theta$
and the limiting measure  as $\nu^\star$.
\end{notn}
% 
%############################################################################
%##############################################################################%##############################################################################
%############################################################################
% #################################################
 \begin{figure}[t]
\centering
\includegraphics[width=3.25in]{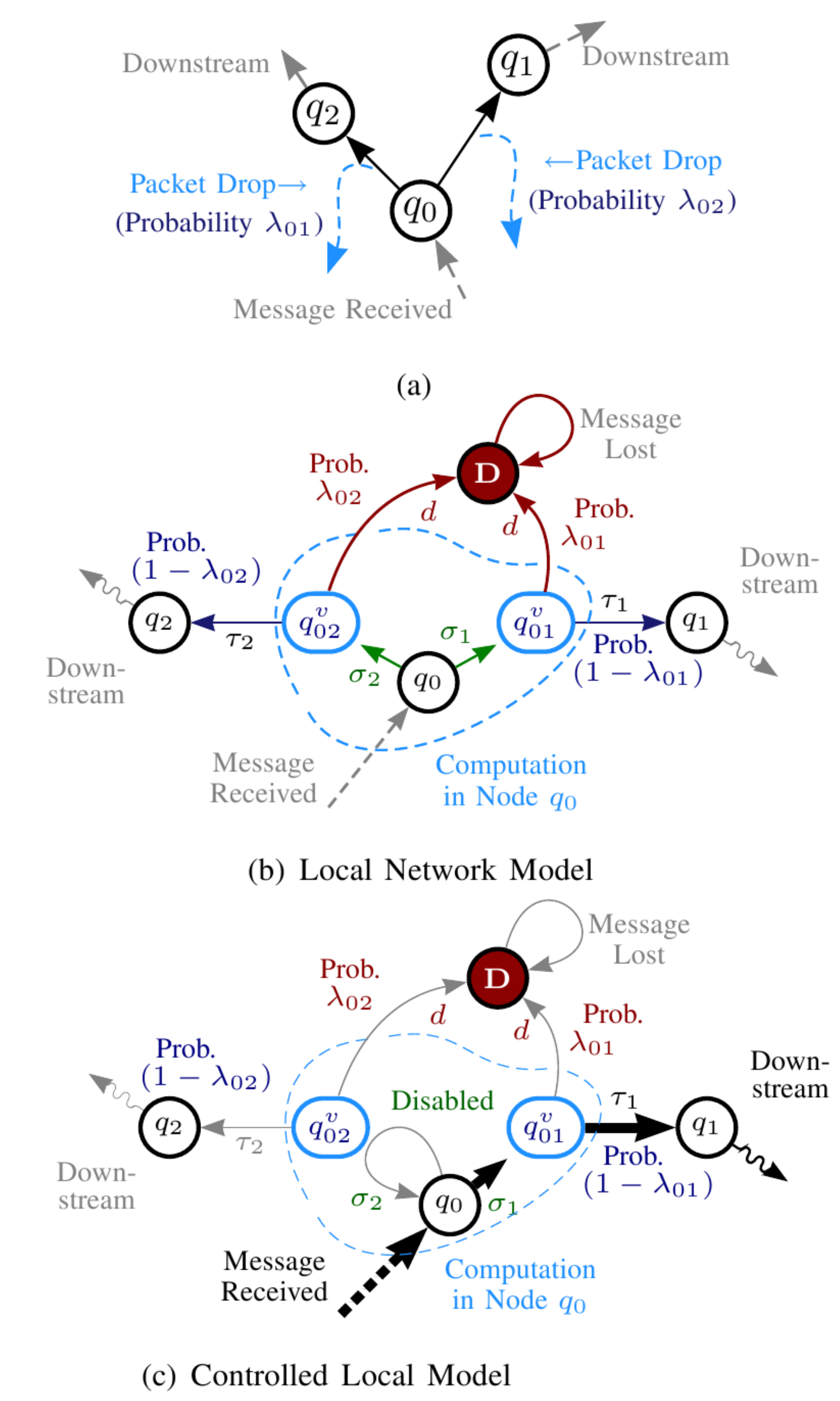}
\caption{Node centric decision for packet forwarding with non-zero drop probability for all choices}\label{figsimT2}
\end{figure}
% #################################################

% \section{Theoretical Development}
\section{Modeling Ad-hoc Networks as PFSA}\label{sec3}
We consider an ad-hoc network of communicating nodes endowed with limited computational resources. For simplicity of exposition, 
we develop the theoretical results under the assumption of a single sink. This is not a serious  restriction and can be easily relaxed.
% {\red More general scenarios will be considered in the later sections.}
The location and identity of the sink is not known a priori to the individual nodes. Inter-node communication links  are assumed to be imperfect, with the possibility of packet drop in each transmission attempt. We  assume  nodes can efficiently gather the following information:
\begin{enumerate}
 \item \textit{(Set of Neighboring Nodes:)} Number and unique id. of nodes to which it can successfully send data via a 1-hop direct link.
% \item A unique network id. for each such neighboring node.
\item \textit{(Local Link Properties:)} Link-specific probability of packet drop for  one-way communication  to a specific neighbor. 
\end{enumerate}
We further assume that the link-specific packet drop probabilities are either constant, or change slowly enough, making  it possible to treat them locally as time-invariant constants for route optimization. Note that this does not imply that the network topology is assumed to be static; we only require that the packet-drop probability for communication from any given node $q_i$ to a particular neighbor $q_j$ be more or less constant, say $0.7$. Thus $q_i$ may choose not to send data to $q_j$ all the time, but when it does, then, on the average, $70 \%$ of the packets get dropped. In practice, the packet drop probabilities may  vary with current network condition, $e.g.$ congestion leading to buffer overflow at specific nodes or (in the context of sensor networks) high-traffic nodes running out of power.  We do not consider these effects in detail; however we briefly describe strategies to handle such effects via simple modifications of the basic principles laid out under the assumption of constant drop probabilities.
 Specific applications, such as wireless sensor networks,  require routing schemes that in addition to  throughput, are aware of  energy and power issues. Also, data-priority need to be  respected  to enable context-aware routing.

First we formalize the modeling of an ad-hoc network as a probabilistic finite state automata.
\begin{defn}[Neighbor Map]\label{defneighbor}
If $Q$ is the set of all nodes in the network, then the neighbor map  $\mathcal{N}:Q \rightarrow 2^Q$ specifies, for each node $q_i \in Q$, the set of nodes $\mathcal{N}(q_i) \subset Q$ (excluding $q_i$) to which $q_i$ can communicate via a single hop direct link.
\end{defn}
\begin{defn}[Packet Drop Probability]\label{defdropprob}
The link specific packet drop probability $\lambda_{ij} \in [0,1]$ is defined to be the limiting ratio of the number of packets dropped to the total number of packets sent, in communicating from node $q_i$ to node $q_j$.
\end{defn}

Note that the drop probabilities are not constrained to be symmetric in general, $i.e.$,   $\lambda_{ij} \neq \lambda_{ji}$. Also, note that we assume the node-based estimation of these ratios to converge fast enough. We  visualize the local network around a node $q_0$ in a manner illustrated in Figure~\ref{figsimT2}(a) (shown for two neighbors $q_1$ and $q_2$). In particular, any packet transmitted from $q_0$ for $q_1$ 
has a drop probability $\lambda_{01}$, and the ones transmitted to $q_2$ have a drop probability $\lambda_{02}$. To correctly represent this information, we require the notion of \textit{virtual nodes} ($q^v_{01},q^v_{02}$ in Figure~\ref{figsimT2}(b)). 

\begin{defn}[Virtual Node]\label{defvirtualnode}
Given a node $q_i$, and a  neighbor $q_j \in \mathcal{N}(q_i)$ with 
a specified drop probability $\lambda_{ij}$, any transmitted data-packet from $q_i$ for $q_j$  is assumed to be first delivered to a virtual node $q^v_{ij}$, upon which there is either an automatic ($i.e.$ uncontrollable) forwarding to  $q_j$ with probability $1-\lambda_{ij}$, or a drop with probability $\lambda_{ij}$. 
The set of all virtual nodes in a network of $Q$ nodes is denoted by $Q^v$ in the sequel.\end{defn}
% 
% \begin{notn}
%  
% \end{notn}
% 
% 
Hence, the total number of virtual nodes is given by:
\begin{gather}
 \Crd(Q^v) = \sum_{i:q_i \in Q} \mathcal{N}(q_i)
\end{gather}
And the cardinality of the set of virtual nodes satisfies:
\begin{gather}\label{eqbnd1}
 0 \leqq \Crd(Q^v) \leqq \Crd(Q)^2 - \Crd(Q)
\end{gather}
% 
% #################################################
% #################################################
\begin{figure}[t]
\centering
% \psfrag{V}[cc]{\footnotesize  \itshape \textshade[0.97]{roundcorners}{\Dblue Virtual Node}}
% \psfrag{E}[cc]{\footnotesize  \itshape \textshade[0.97]{roundcorners}{\Dgreen Controllable}}
% \psfrag{F}[bc]{\footnotesize  \itshape \textshade[0.97]{roundcorners}{Uncontrollable}}
% \psfrag{N}[tc]{\footnotesize \bf A 6-node Network}
% \psfrag{D}[cr]{\footnotesize  \BRed \itshape \textshade[0.97]{roundcorners}{Dump State}}
% \psfrag{M}[ct]{\footnotesize \bf PFSA model with virtual nodes}
% \psfrag{q}[cc]{\footnotesize \txt{$q_1$}}
% \psfrag{SD}[cc]{\BRed\footnotesize \txt{$\lambda_{12}$}}
% \psfrag{m}[cc]{\footnotesize \txt{$q_2$}}
% \psfrag{v}[cr]{\Nblue\footnotesize \txt{$q^v_{12}$}}
 \includegraphics[width=3.25in]{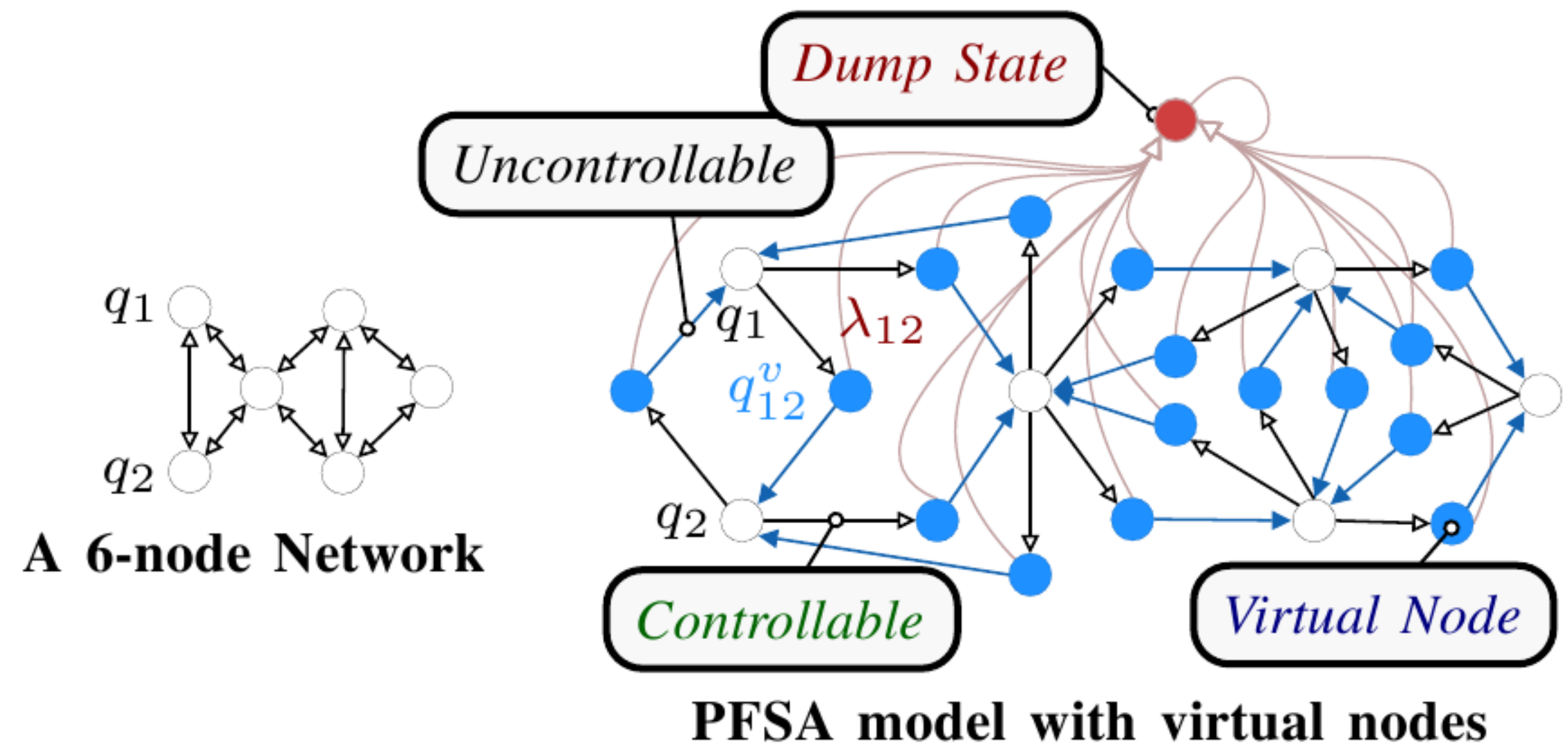}
\caption{6-node Network and  PFSA model with 23 states (16 virtual nodes, 6 nodes, 1 dump state)}\label{fignetpfsa}
\end{figure}
% #################################################
% #################################################
We are  ready to  model  an ad-hoc network as a PFSA.
\begin{defn}[PFSA Model of Network]\label{defpfsanetwork}
 For a given set of nodes $Q$, the function $\mathcal{N}:Q \rightarrow 2^Q$,  the link specific drop probabilities $\lambda_{ij}$ for any node $q_i$ and a neighbor $q_j \in \mathcal{N}(q_i)$, and a specified sink $q_\Sn \in Q$, the PFSA $\Gnet = (Q^N,\Sigma,\delta,\widetilde{\Pi},\chi,\mathscr{C})$ is defined to be a model of the network, where (denoting $\Crd(\mathcal{N}(q_i))=m$):
\begin{subequations}
\begin{align*}
&\textrm{\scshape \sffamily \footnotesize $\circ$ States:}
% \notag\\
%  &
\mspace{50mu}Q^N = Q \bigcup Q^v \bigcup \big \{q_D\big \}
\intertext{where $Q^v$ is the set of virtual nodes, and $q_D$ is a dump state which models packet loss. For the alphabet $\Sigma$:}
% \vspace{-10pt}
% \intertext{
&\textrm{\scshape \sffamily \footnotesize $\circ$ Alphabet:}
% \notag\\
% &
\mspace{50mu}\Sigma = \bigcup_{i:q_i \in Q} \left ( \bigcup_{j: q_j \in \mathcal{N}(q_i)} \sigma_{ij} \right )\bigcup \big \{\sigma_D\big \} 
\intertext{$\sigma_{ij}$ denotes  transmission (attempted or actual)
from $q_i$ to $q_j$, and $\sigma_D$ denotes transmission to  $q_D$
(packet loss).}
&\txt{\scshape \sffamily \footnotesize $\circ$ Transition \\\scshape \sffamily \footnotesize Map: $\phantom{XXXX.X.}$} \ 
% \notag\\
% &\mspace{50mu}
\mspace{5mu}
\delta(q,\sigma) = \left \{ \begin{array}{cl}
       q^v_{ij} & \textrm{if } q = q_i , \sigma = \sigma_{ij} \\
q_j & \textrm{if } q = q^v_{ij},\sigma = \sigma_{ij}\\
q_D & \textrm{if } q = q^v_{ij},\sigma = \sigma_D \\
q_D & \textrm{if } q = q_D,\sigma = \sigma_D \\
- & \textrm{undefined otherwise}
                            \end{array}
\right.\\
&
% \mspace{-24mu}
\txt{\scshape \sffamily \footnotesize $\circ$ Probability \\ \scshape \sffamily \footnotesize Morph $\phantom{XXXXx}$\\\scshape \sffamily \footnotesize  Matrix:$\phantom{XXXXx}$}
%\notag\\
%&
\mspace{3mu}
\ \widetilde{\Pi}(q,\sigma) = \left \{ \begin{array}{cl}
       \frac{1}{m} & \textrm{if } q = q_i , \sigma = \sigma_{ij} \\
1-\lambda_{ij} & \textrm{if } q = q^v_{ij},\sigma = \sigma_{ij}\\
\lambda_{ij} & \textrm{if } q = q^v_{ij},\sigma = \sigma_D \\
1 & \textrm{if } q = q_D,\sigma = \sigma_D \\
0 & \textrm{otherwise}
                            \end{array}
\right.\\
&\txt{\scshape \sffamily \footnotesize $\circ$ Characteristic \\ \scshape \sffamily \footnotesize  Weights:$\phantom{XXXXXX..}$}
% \notag\\
% &
\mspace{50mu}\chi_i = \left \{ \begin{array}{cl}
       1 & \textrm{if } q_i = q_\Sn            \\
0 & \textrm{otherwise}
                  \end{array}
\right.\\
&\txt{\scshape \sffamily \footnotesize $\circ$  Controllable \\\scshape \sffamily \footnotesize Transitions:$\phantom{Xxx}$}
% \notag\\
% &
\mspace{15mu}
\ \forall q_i \in Q, q_j \in \mathcal{N}(q_i), q_i \xrightarrow{\sigma_{ij}} q^v_{ij} \in \mathscr{C}
\end{align*}
\end{subequations}
\end{defn}
We note  that for a network of $Q$ nodes, the PFSA model may have (almost always has, see Figure~\ref{fignetpfsa}) a significantly larger number of states. Using Eq.~\eqref{eqbnd1}:
\begin{gather}
 \Crd(Q^N) = \Crd(Q) + \Crd(Q^v) + 1 \\
\Rightarrow \Crd(Q) + 1 \leqq \Crd(Q^N) \leqq \Crd(Q)^2 + 1
\end{gather}
This state-explosion will not be a problem for the distributed approach developed in the sequel, since we use the complete model $\Gnet$ only for the purpose of deriving theoretical guarantees.
% and as a performance benchmark for the proposed distributed routing algorithms.
% 
Note, that Definition~\ref{defpfsanetwork} generates a PFSA model which can be optimized in a straightforward manner using the language-measure-theoretic technique described in Section~\ref{sec2} (See \cite{CR07}) for details). This would yield the optimal routing policy in terms of the disabling decisions at each node that minimize source-to-sink drop probabilities (from every node in the network).   To see this explicitly, note that the measure-theoretic approach elementwise maximizes 
% \begin{gather}
 $\lim_{\theta \rightarrow 0^+}  \theta \big [ \mathbb{I} - (1-\theta)\Pi \big ]^{-1} \chi = \Q \chi$, 
% \end{gather}
where the $i^{th}$ row of $\Q$ (denoted as $\wp^i$) is the stationary probability vector for the PFSA initialized at state $q_i$ (See Proposition~\ref{prop1}). Since, the dump state has characteristic $-1$, the sink has characteristic $1$, and all other nodes have characteristic $0$, it follows that this optimization maximizes the quantity $\wp^i_\Sn - \wp^i_{\textrm{\scshape Dump}}$, for every source state or node $q_i$ in the network. Note that $\wp^i_\Sn, \wp^i_{\textrm{\scshape Dump}}$ are the stationary probabilities of reaching the sink and incurring a packet loss to dump respectively, from a given source $q_i$.
Thus, maximizing $\wp^i_\Sn - \wp^i_{\textrm{\scshape Dump}}$ for every $q_i \in Q$ guarantees that the computed routing policy is indeed optimal in the stated sense.
However, the procedure in \cite{CR07}  requires  centralized computations, which is precisely what we wish to avoid. 
The key technical contribution in this paper is to  develop a distributed approach to language-measure-theoretic PFSA optimization. In effect, the theoretical development in the next section allows us to carry out the language-measure-theoretic optimization of a given PFSA, in situations where we do not have access to the complete $\Pi$ matrix, or the $\chi$ vector at any particular node ($i.e.$ each node has a limited local view of the network), and are restricted to communicate only with immediate neighbors.
We are interested in not just computing the measure vector in a distributed manner, but 
optimizing the PFSA via selected disabling of controllable transitions (See Section~\ref{sec2}).
This is  accomplished by Algorithm~\ref{AlgorithmOPT}.

Before we embark up on the detailed analysis of Algorithm~\ref{AlgorithmOPT} in the next section, we briefly elucidate the connection with decentralized Markov Decision Processes. The PFSA based modeling framework is somewhat different from the standard MDP architecture. For example, in contrast to the latter,  our actions are "controllable" transitions, and have  probabilities associated with them. 
Rewards and penalties are not associated with individual actions, but with state visitations (and modeled via the characteristic weights). We maximize the long term or expected reward by maximizing the probability of reaching the sink, while simultaneously minimizing the probability of reaching the dump state, $i.e.$, a drop, from any arbitrary node in the network. More details on relations to the standard approach is given in \cite{CR10}.
\section{Decentralized PFSA Optimization}\label{sec4}
% #################################################
% ##############################################################
% ##############################################################
% ##############################################################
% ##############################################################
\setlength{\algomargin}{0em} 
\begin{algorithm}[t]
\small
 \SetLine
\linesnotnumbered
\dontprintsemicolon
  \SetKwInOut{Input}{input}
  \SetKwInOut{Output}{output}
 \SetKw{Tr}{true}
   \SetKw{Tf}{false}
  \caption{{\small  Distributed Update of Node Measures}}\label{AlgorithmOPT}
\Input{$\Gnet=(Q,\Sigma,\delta,\widetilde{\Pi},\chi,\mathscr{C})$, $\theta$}
% \Output{Updated measure values}
%
\Begin{
Initialize $\forall q_i \in Q, \widehat{\nu}^\cur_\theta \vert_i = 0$\;
\txt{\sffamily \footnotesize \BRed $/* \ $ {Begin Infinite \uline{Asynchronous} Loop} $\ */ $ } 
\While{\Tr}{
\For{each node $q_i \in Q$}{
\If{ $\mathcal{N}(q_i) \neq \varnothing $}{

$m = \Crd(\mathcal{N}(q_i))$\;
\For {each node $q_j \in \mathcal{N}(q_i)$}{
\BlankLine
\Nblue
\txt{\sffamily \footnotesize \color{DodgerBlue2} $/* \ $ { (a1) Internode Communication} $\ */ $ \color{black}\\ $\phantom{.}$}%
\vspace{-5pt}
\colorbox{LightCyan}{
  \Mblue Query $\widehat{\nu}^\cur_\theta \vert_j$ \& Drop Prob. $\lambda_{ij}$\;} \color{black}
\BlankLine
\BlankLine
\txt{\sffamily \footnotesize \color{IndianRed4} $/* \ $ { (a2) Control Adaptation} $\ */ $ \color{black}}%
% \vspace{-8pt}
% \parashade[.99]{sharpcorners}
\colorbox{MistyRose}{
\begin{minipage}{2in}
\eIf{$ \widehat{\nu}^\cur_\theta \vert_j < \widehat{\nu}^\cur_\theta \vert_i$}{
% 
% \Dgreen
$\Pi_{ii} = \Pi_{ii} + \Pi_{i(q^V_{ij})}$\;
$\Pi_{i(q^V_{ij})} = 0$\color{black}\tcc*[r]{\Dgreen  \footnotesize  {\bf \sffamily   Disable} \color{black}}
} {
\If{ $\Pi_{i(q^V_{ij})} == 0$}{
% \Dgreen
$\Pi_{i(q^V_{ij})} = \frac{1}{m}$\;
$\Pi_{ii} = \Pi_{ii} - \frac{1}{m}$ \color{black}\txt{\sffamily \footnotesize \Dgreen $/* \ $ {\bf \sffamily Enable} $\ */ $ \color{black}}
}
}
\end{minipage}
}
\vspace{2pt}
\BlankLine
\txt{ \sffamily \footnotesize \color{Gold4} $/* \ $ { (a3) Updating Virtual Nodes} $\ */ $ \color{black}\\ $\phantom{.}$} % 
\vspace{-5pt}
\Mblue
\colorbox{Beige}{
$\widehat{\nu}^\cur_\theta \vert_{(q_{ij}^V)} = (1-\theta)(1 - \lambda_{ij}) \widehat{\nu}^\cur_\theta \vert_j$}\color{black}
}
}
\txt{ \sffamily \footnotesize \color{Chartreuse4} $/* \ $ { (a4) Updating  Node} $\ */ $ \color{black}\\ $\phantom{.}$} % 
\vspace{-5pt}
\BRed
% \vspace{-16pt}
% \parashade[.99]{roundcorners}
\colorbox{Honeydew}{
% \vspace{-10pt}
% \begin{multline*}
\txt{
$
\widehat{\nu}^\cur_\theta \vert_i = \displaystyle\mspace{-25mu} \sum_{j: q_j \in \mathcal{N}(q_i)} \mspace{-20mu}(1 -\theta) \Pi_{i(q^V_{ij})} \widehat{\nu}^\cur_\theta \vert_{(q_{ij}^V)} $\\ \hspace{80pt}$+ (1-\theta)\Pi_{ii} \widehat{\nu}^\cur_\theta \vert_i + \theta\chi \vert_i
$}%  \end{multline*}
% \vspace{-10pt}
} 
% \vspace{-10pt}
\color{black}
}}
}
\end{algorithm}
% ##############################################################
% ##############################################################
% ##############################################################
\begin{notn}
 In the sequel, the current measure value, for a given $\theta$, at node $q_i \in Q$ is denoted as $\widehat{\nu}^\cur_\theta \vert_i$, and the measure of the virtual node $q^v_{ij} \in Q^N$ is denoted as $\widehat{\nu}^\cur_\theta \vert_{(q_{ij}^V)}$. The parenthesized entry $(q_{ij}^V)$ denotes the index  of the virtual node $q^v_{ij}$ in the state set $Q^N$. Similarly, the 
transition probability from $q_i$ to $q^v_{ij}$ is denoted as $\Pi_{i(q_{ij}^V)}$. The subscript entry $i(q_{ij}^V)$ denotes the 
$ik^{th}$ element of $\Pi$, where $k = (q_{ij}^V)$.
\end{notn}

Algorithm~\ref{AlgorithmOPT} establishes a distributed, asynchronous update procedure which 
achieves the following:
% \begin{gather}
\cgather[5pt]{ \forall q_i \in Q, \widehat{\nu}_\theta \vert_i  \xrightarrow[\txt{\footnotesize convergence}]{\txt{\footnotesize global }} \nu^\star_\theta \vert_i}
% \end{gather}
where $\nu^\star_\theta \vert_i$ is the optimal measure for $q_i \in Q$ that would be obtained by optimizing the PFSA $\Gnet$, for a given $\theta$, in a centralized approach (See Section~\ref{sec2}).
The optimal routing policy can then be obtained by forwarding packets to neighboring nodes which 
have a better or equal current measure value. If more than a one such neighbor is available, then one chooses the forwarding node randomly, in an equiprobable manner. 
In fact, the nodes need not wait for exact convergence; in the sequel we show that this forwarding policy  converges to the  globally optimal 
routing policy, that, for a sufficiently small $\theta$, it maximizes probability of reaching the sink, while 
simultaneously minimizing the probability of packet drops. Furthermore, choosing randomly between qualifying neighboring nodes leads to 
significant congestion resilience. These issues would be elaborated in the sequel (Proposition~\ref{propchar}). First,  Algorithm~\ref{AlgorithmOPT} is analyzed to establish convergence.

Algorithm~\ref{AlgorithmOPT} has four distinct parts, marked as (a1), (a2), (a3) and (a4). Part (a1) involves 
internode communication, to enable a particular node $q_i \in Q$ to ascertain the current measure values of neighboring nodes, and the 
drop probabilities $\lambda_{ij}$ on respective links. Recall, that we assume the probabilities $\lambda_{ij}$ to be more or less constant; nevertheless nodes 
estimate these values to adapt to changing (albeit slowly) network conditions. Part (a2) is the control adaptation, in which the nodes decide, based on local information, the set of forwarding nodes. Part (a3) is the computation of the updated measure values for the virtual nodes $q^v_{ij}$ where $j: q_j \in \mathcal{N}(q_i)$. Finally, part (a4) updates the
measure of the node $q_i$ based on the computed current measures of the virtual nodes. We note that Algorithm~\ref{AlgorithmOPT} only uses information that 
is either available locally, or that which can be queried from neighboring nodes.

\begin{prop}[Convergence]\label{propmain}
For a  network $Q$ modeled as a PFSA $\Gnet=(Q^N,\Sigma,\delta,\widetilde{\Pi},\chi, \mathscr{C})$, the distributed procedure 
in Algorithm~\ref{AlgorithmOPT} has the following properties:
 \begin{enumerate}
  \item Computed measure values for every node $q_i \in Q$ are non-negative and bounded above by $1$, $i.e.$,
\begin{gather}
 \forall q_i \in Q^N, \forall t \in [0,\infty), \ \widehat{\nu}^t_\theta \vert_i \in [0, 1]
\end{gather}
\item For constant drop probabilities and constant neighbor map $\mathcal{N}:Q \rightarrow 2^Q$, Algorithm~\ref{AlgorithmOPT} converges in the sense:
\begin{gather}
 \forall q_i \in Q^N, \lim_{t \rightarrow \infty}\widehat{\nu}^t_\theta \vert_i =  \nu^\infty_\theta \vert_i \in [0,1]
\end{gather}
\item Convergent measure values coincide with the optimal values computed by the centralized approach:
\begin{gather}
 \forall q_i \in Q^N, \nu^\infty_\theta \vert_i =  \nu^\star_\theta \vert_i
\end{gather}

 \end{enumerate}

\end{prop}
\begin{IEEEproof}
\textit{(Statement 1:)} Non-negativity of the measure values is obvious. 
For establishing the upper bound, we use induction on computation time $t$. 
We note that all the measure values $\widehat{\nu}^t_\theta \vert_i$ are initialized to $0$ at time $t=0$.
The first node to change its measure will be the sink, which is updated at some time $t=t_0$:
\begin{gather}\label{eqt0}
 \widehat{\nu}^{t_0}_\theta \vert_{(q_\Sn)} = 0 + \theta\chi_{(q_\Sn)} = \theta
\end{gather}
where the first term is zero since all nodes still have measure zero and the sink  characteristic $\chi_{(q_\Sn)} = 1$.
Thus,  there exists a non-trivial time instant $t_0$, at which:
\begin{align}
\textrm{\sffamily {\small (Induction Basis)} }  \forall q_i \in Q^N,  \widehat{\nu}^t_\theta \vert_i \leqq 1
\end{align}
% At initialization, $i.e.$, at time $t=0$, we have 
% \begin{align}
%  \textrm{\sffamily\textsc{(Induction Basis)}} &&& \widehat{\nu}^0_\theta \vert_i =\left \{ \begin{array}{ll}
%        	                          1, & \mathrm{if} \ q_i = q_\Sn \\
% 				  0, & \mathrm{otherwise}
%                                           \end{array}\right.
% \end{align}
% 
Next we assume for  time $t=t'$, we have
\begin{align*}
\textrm{\sffamily {\small (Induction Hypothesis)} }  \forall q_i \in Q^N, \forall \tau \leqq t', \widehat{\nu}^\tau_\theta \vert_i \leqq 1
\end{align*}
We consider the next updates for physical nodes and virtual nodes separately, and denote the time instant for the next updates as $t'_+$. Note, that 
$t'_+$ actually may be different for different nodes (asynchronous operation).\\
{\itshape (Virtual Nodes)}
For any virtual node $q_i  = q^v_{kj} \in Q^N$, where $q_k,q_j \in Q$, we have:
\begin{gather}
 \widehat{\nu}^{t'_+}_\theta \vert_i = (1-\lambda_{ij})(1-\theta)\widehat{\nu}^{t'}_\theta \vert_j \leqq 1
\end{gather}
{\itshape (Physical Nodes)}
For any  $q_i   \in Q$, where  set of enabled neighbors
$E_n = \big \{q_j \in \mathcal{N}(q_i) \ \textrm{s.t. } \widehat{\nu}^{t'_+}_\theta \vert_{(q_{ij}^V)} \geqq \widehat{\nu}^{t'}_\theta \vert_i\big \}$:
\begin{multline*}
 \widehat{\nu}^{\tau+}_\theta \vert_i =\frac{1}{\Crd(\mathcal{N}(q_i))}\bigg (\sum_{j:q_j \in E_n} (1-\theta)^2(1-\lambda_{ij})\widehat{\nu}^{t'}_\theta \vert_{(q_{ij}^V)} \\+ \sum_{j: j \in \mathcal{N}(q_i) \setminus E_n}(1-\theta)\widehat{\nu}^{t'}_\theta \vert_i \bigg ) \\\leqq
\frac{1}{\Crd(\mathcal{N}(q_i))}\bigg (\sum_{j:q_j \in E_n} 1 + \sum_{j: j \in \mathcal{N}(q_i) \setminus E_n}1 \bigg )
\leqq  1
\end{multline*}
which establishes Statement 1.\\
\textit{(Statement 2:)}
We claim that for each node $q_i \in Q^N$, the sequence of measures $\widehat{\nu}^t_\theta\vert_i$ forms a monotonically non-decreasing sequence as a function of the computation time $t$. Again, we use induction on computation time. 
% 
% Algorithm~\ref{AlgorithmOPT} induces a monotonically non-decreasing sequence of measures,
% 
Considering the time instant $t_0$ (See Eqn.~\eqref{eqt0}), we note that we have an instant up to which all measure values have indeed changed in a non-decreasing fashion, since the measure 
of $q_\Sn$ increased to $\theta$, while  other nodes are still at $0$; which establishes the  basis.
For our  hypothesis, we assume that there exists  some time instant $t' > t_0$, such that all measure values  have undergone non-decreasing updates up to $t'$.
% \begin{align}
% \mspace{-13mu}\textrm{\sffamily\textsc{(Induction Hypothesis)}} &&& \forall q_i \in Q^N, \forall \tau \leqq t', \widehat{\nu}^\tau_\theta \vert_i \geqq \widehat{\nu}^{t_0}_\theta \vert_i
% \end{align}
We consider the physical node $q_i \in Q$ which is the first one to update next, say at the instant $t'_+ > t'$. Referring to Algorithm~\ref{AlgorithmOPT}, this update occurs by first updating the set of virtual nodes $\{q_{ij}^v : q_j \in \mathcal{N}(q_i)\}$. Since  virtual nodes update as:
\begin{gather}
 \widehat{\nu}^{t'_+}_\theta \vert_{(q^v_{ij}} = (1-\theta)(1-\lambda_{ij})\widehat{\nu}^{t'}_\theta \vert_j
\end{gather}
it follows from the induction hypothesis that 
\begin{gather}
 \widehat{\nu}^{t'_+}_\theta \vert_{(q^v_{ij})} \geqq \widehat{\nu}^{t'}_\theta \vert_{(q^v_{ij})}
\end{gather}
% 
% 
% We consider the next update instant  for the
% node $q_i$, which is the first one to update after
%  physical node $q_i \in Q$.
 If the connectivity ($i.e.$ the forwarding decisions) for the physical node $q_i$  remains unchanged for the instants $t'$ and $t'_+$, and since 
the measures of any neighboring node has not decreased (by induction hypothesis), then:
\begin{gather}
 \widehat{\nu}^{t'_+}_\theta \vert_i \geqq \widehat{\nu}^{t'}_\theta \vert_i
\end{gather}
If, on the other hand, the set of disabled transitions for $q_i$ changes ($e.g.$ for some $q_j \in \mathcal{N}(q_i)$, $q_i \xrightarrow{\sigma_{ij}}q_{ij}^v$ was disabled at $t'$ and is enabled at $t'_+$, or vice verse), the measure of node $q_i$ is increased by the additive factor $\frac{(1-\theta)}{\Crd(\mathcal{N}(q_i))}\bigg\vert\widehat{\nu}^{t'}_\theta \vert_i - \widehat{\nu}^{t'}_\theta \vert_{(q^v_{ij})} \bigg \vert  $, which completes the inductive process and establishes our claim that the measure values form a non-decreasing sequence for each node as a function of the computation time. Since, a non-decreasing bounded sequence in a complete space must converge to a unique limit~\cite{R88},  the convergence: 
\begin{gather}
 \forall q_i \in Q^N, \lim_{t \rightarrow \infty}\widehat{\nu}^t_\theta \vert_i =  \nu^\infty_\theta \vert_i \in [0,1]
\end{gather}
  follows from the 
existence of the upper bound established in Statement 1. This establishes Statement 2.\\
\textit{(Statement 3:)}
From the update equations in Algorithm~\ref{AlgorithmOPT}, we note that the limiting measure values satisfy:
\begin{align}
\widehat{\nu}_\theta^{\infty} \big \vert_i &= (1-\theta)\sum_{j \in \mathcal{N}(i)} \Pi_{ij} \widehat{\nu}_\theta^{\infty} \big \vert_j + \theta \chi \vert_i \notag\\
\Rightarrow \widehat{\nu}_\theta^{\infty} &= \theta \big [ \mathbb{I} - (1-\theta)\Pi \big ]^{-1}\chi
\end{align}
which implies that measure values does indeed converge to the measure vector computed in a centralized fashion (See Eq.~\eqref{eqmesd}).
Noting that any further disabling (or re-enabling) would not increase the measure values computed by Algorithm~\ref{AlgorithmOPT}, we conclude that this must be 
the optimal disabling set that would be obtained by the centralized language-measure theoretic optimization of  PFSA $\Gnet$ (Section~\ref{sec2}). This completes the proof.
\end{IEEEproof}
\begin{prop}[Initialization Independence]\label{propinitindependence}
For a  network $Q$ modeled as a PFSA $\Gnet=(Q^N,\Sigma,\delta,\widetilde{\Pi},\chi, \mathscr{C})$, convergence of Algorithm~\ref{AlgorithmOPT} is independent of the initialization of the measure values, $i.e.$, if $\widehat{\nu}_{\theta,\alpha}^t$  denotes  the measure vector at time $t$ with arbitrary initialization  $\alpha \in [0,1]^{\Crd(Q^N)}$, then:
\begin{gather}\label{eqprop1}
 \lim_{t\rightarrow \infty}\widehat{\nu}_{\theta,\alpha}^t = \lim_{t\rightarrow \infty}\widehat{\nu}_{\theta}^t
\end{gather}
% \begin{gather}
%  \widehat{\nu}_{\theta,\alpha}^0= \alpha , \widehat{\nu}_{\theta}^0= [0 \cdots 0]^T
% \end{gather}
% 
where $\widehat{\nu}_{\theta,\alpha}^0= \alpha$ and $ \widehat{\nu}_{\theta}^0= [0 \cdots 0]^T$.
% (as before) is the measure vector at time $t$ with all nodes initialized to zero.
\end{prop}
\begin{IEEEproof} The measure update equations in Algorithm~\ref{AlgorithmOPT} dictate that the measure values will have a positive contribution from $\alpha$. Denoting the contribution of $\alpha$ to the measure of node $q_i \in Q$ at time $t$ as $\mathcal{C}^t_\alpha(q_i)$, we note that the measure can be written as $\widehat{\nu}_{\theta,\alpha}^t = \mathcal{C}^t_\alpha(q_i) + f_i^t$, where $f_i^t$ is independent of $\alpha$.
Furthermore, the linearity of the updates imply that $\mathcal{C}^t_\alpha(q_i)$ can be used to formulate an inductive argument as follows. We use $k_\star^t \in \mathbb{N}\cup \{0\}$  to denote the minimum number of updates  that every node in the network has encountered up to  time instant $t \in [0,\infty)$. 
% , and 
% $k_{i}^t \in \mathbb{N}\cup \{0\}$ denotes the number of updates for the specific physical node $q_i \in Q$ up to time $t$, implying that $ k_\star^t  \leqq k_i^t $. 
We claim that:
\begin{gather}\label{eqclaim1}
 \forall q_i \in Q, \forall t \in [0,\infty), \mathcal{C}^t_\alpha(q_i) \leqq (1-\theta)^{k_\star^t} \vert \vert \alpha \vert \vert_1
\end{gather}
To establish this claim, we use induction on $k_\star^t$.
For the  basis, we note that there exists a time instant $t_0$, such that $\forall \tau \leqq t_0, k_\star^\tau =0$, implying that 
\begin{gather*}
 \forall \tau \leqq t_0, \mathcal{C}^\tau_\alpha(q_i) = \alpha_i \leqq (1-\theta)^0 \sum_{q_j \in Q} \alpha_j = (1-\theta)^{k_\star^\tau} \vert \vert \alpha \vert \vert_1
\end{gather*}
We assume that if at some  $t_k$,  $k_\star^{t_k} = k \in \mathbb{N}$, then:
\begin{align*}
%  \mspace{-13mu}
\textrm{\sffamily {\small (Induction Hypothesis)} }  \forall q_i \in Q, \mathcal{C}^{t_k}_\alpha(q_i) \leqq (1-\theta)^k \vert \vert \alpha \vert \vert_1
\end{align*}
Next let $q_i\in Q$ be an arbitrary physical node, and we consider the first update of $q_i$ at $t_k^+ > t_k$:
\begin{multline*}
\widehat{\nu}^{t_k^+}_\theta \vert_i = \mspace{-25mu} \sum_{j: q_j \in \mathcal{N}(q_i)} \mspace{-20mu}(1 -\theta) \Pi_{i(q^v_{ij})} \widehat{\nu}^{t_k}_\theta \vert_{(q_{ij}^v)} + (1-\theta)\Pi_{ii} \widehat{\nu}^{t_k}_\theta \vert_i + \theta\chi_i \\
\Rightarrow \mathcal{C}^{t_k^+}_\alpha (q_i)\leqq  \mspace{-28mu}\sum_{j: q_j \in \mathcal{N}(q_i)} \mspace{-20mu}(1 -\theta) 
\Pi_{i(q^v_{ij})} (1-\lambda_{ij})(1-\theta)(1-\theta)^k\vert \vert \alpha \vert \vert_1  \\ + (1-\theta)\Pi_{ii}(1-\theta)^k\vert \vert \alpha \vert \vert_1 + \theta\chi_i\\
% \end{multline}
% 
% \begin{gather}
\Rightarrow \mathcal{C}^{t_k^+}_\alpha (q_i)\leqq (1-\theta)^{k+1}\vert \vert \alpha \vert \vert_1 \mspace{325mu}
\end{multline*}
We note that if $k_\star^{t_{k+1}} = k+1$, then every node $q_i \in Q$ must have undergone one more update 
since  $t_k$ implying:
\begin{gather}
\forall q_i \in Q, \mathcal{C}^{t_{k+1}}_\alpha (q_i)\leqq (1-\theta)^{k+1}\vert \vert \alpha \vert \vert_1
\end{gather}
which completes the induction proving  Eq.~\eqref{eqclaim1}.
Observing that $\lim_{t\rightarrow\infty} k_\star^t = \infty$, and  $\vert \vert \alpha \vert \vert_1 < \infty$, we conclude:
\begin{gather}
 \forall q_i \in Q, \lim_{t\rightarrow\infty} \mathcal{C}^t_\alpha(q_i) =0
\end{gather}
which immediately implies Eq.~\eqref{eqprop1}.
\end{IEEEproof}
Next we investigate the performance of the proposed approach, and establish guarantees on global performance achieved via  local
decisions dictated by Algorithm~\ref{AlgorithmOPT}.
We need some technical lemmas, and the notion of 
strongly absorbing graphs, and graph powers. 
\begin{defn}[Exact Power of Graph]
For a given graph $G=(V,E)$, the exact power $G^d$, for $d \in \mathbb{N}$, is a graph $(V,E')$, such that 
 $(q_i,q_j)$ is an edge in $G^d$, only if there exists a sequence of edges of length exactly $d$ from node $q_i$ to node $q_j$ in $G$.
 \end{defn}
\begin{defn}[Strongly Absorbing Graph]\label{defNSA}
 A finite  directed graph $G=(V,E)$ ($V$ is the set of nodes and $E \subseteq V \times V$  the set of edges) is defined to be  strongly absorbing (SA), if:
\begin{enumerate}
 \item There are one or more absorbing nodes, $i.e.$, $\exists A \subsetneqq V$, s.t. every node in $A$ (non-empty) is absorbing.
\item There exists at least one sequence of edges from any node to one of the absorbing nodes in $A$.
\item If $E^d$ denotes the set of edges for the $d^{th}$ exact power of $G$, then, for distinct nodes $q_i,q_j \in V$, 
\begin{gather}
 (q_i,q_j) \in E \Rightarrow  \forall d \in \mathbb{N}, \ (q_j,q_i) \notin E^d
\end{gather}
% \item For every non-absorbing node, we have:
% \begin{gather}
%        \forall q_i \in V\setminus A, \exists q_j \in V \ \textrm{s.t.} \  (q_j,q_i) \notin E
%       \end{gather}
% 
\end{enumerate}

\end{defn}
% #################################################
% #################################################
% #################################################
% \begin{figure}[t]
% \centering
% \VCDraw[.7]{%
% \begin{VCPicture}{(-2,-.5)(6,3)}
% \State[q_3]{(0,0)}{A} 
% \State[q_4]{(2,0)}{B} 
% \State[q_5]{(4,0)}{C} 
% \State[q_1]{(-2,2.2)}{D} 
% % \SetEdgeLineColor{DarkRed}
% \State[q_2]{(0,2.2)}{F} 
% \SetStateLineColor{DarkRed}
% \State[\BRed b_1]{(2,2.4)}{E} 
% \State[\BRed b_2]{(6,2)}{G} 
% \EdgeL[0.8]{A}{B}{}
% \EdgeL[0.8]{B}{C}{}
% \EdgeL[0.8]{C}{G}{}
% \EdgeL[0.8]{C}{E}{}
% \EdgeL[0.8]{A}{E}{}
% \EdgeL[0.8]{D}{A}{}
% \EdgeL[0.8]{D}{F}{}
% \EdgeL[0.8]{F}{E}{}
% \LArcL[0.8]{F}{G}{}
% \LoopE[.7]{G}{}
% \LoopE[.7]{E}{}
% \LoopW[.7]{D}{}
% % \LoopN[.7]{B}{}
% \end{VCPicture}}
% \caption{SA graph (absorbing set $A=\{b_1,b_2\}$)}\label{figstrongabs}
% \end{figure}
% % #################################################
% % #################################################
% % #################################################
% 
\begin{lem}[Properties of SA Graphs]\label{lemNSA}
Given a SA graph $G=(V,E)$, with $A \subsetneqq V$ the absorbing set:
\begin{enumerate}
\item The  power graph $G^d$ is SA for every $d \in \mathbb{N}$.
 \item $ q \notin A \Rightarrow  \exists q' \in V\setminus \{ q\} \ \textrm{s.t.} \ (q',q) \notin E$ 
\item $\exists d \in \mathbb{N} \left (  \forall q \in V\setminus A \left ( \exists q' \in A \left ( (q,q') \in E^d \right ) \right ) \right )$
\end{enumerate}
\end{lem}
\begin{IEEEproof} Statement 1 is immediate from Definition~\ref{defNSA}.
 Statement 2 follows immediately from noting:
\begin{gather*}
 q \notin A \Rightarrow \exists q'\in V\setminus \{ q\} \ \textrm{s.t.} \ (q,q') \in E \Rightarrow (q', q) \notin E
\end{gather*}
Statement 3 follows, since from each node there is a path (length bounded by $\Crd(V)$) to a absorbing state.
\end{IEEEproof}
% 
% We will show in the sequel, that probabilistic machines with strongly absorbing graphs have the largest non-unity eigen value of the induced Markov chain bounded above by 
% $1 - p_\ell$ where $p_\ell$ is the maximum self-loop probability. This result will be crucial in establishing the convergence and $\epsilon$-optimality  of the 
% proposed algorithm. First, 
% 
% We observe that the optimized network is guaranteed to be a PFSA with a strongly absorbing graph.
The performance of such control policies, and particularly the convergence time-complexity is closely related to the 
spectral gap of the induced Markov Chains. Hence we need to compute lower bounds on the spectral gap of the chains arising in the context of the proposed optimization, which (as we shall see later) have the 
strongly absorbing property. The following result  computes such a bound as a simple function of the non-unity diagonal entries of $\Pi$.
\begin{prop}[Spectral Bound]\label{propspectral}
 Given a n-state PFSA $G=(Q,\Sigma,\delta,\widetilde{\Pi})$ with a  strongly absorbing graph, the magnitude of non-unity eigenvalues of the transition  matrix $\Pi$ is bounded above by the maximum non-unity diagonal entry of $\Pi$.
\end{prop}
\begin{IEEEproof}
Without loss of generality, we assume that $G$ has a single absorbing state (distinct absorbing states can be merged without affecting non-unity eigenvalues). Now, $\mu$ is an eigenvalue of $\Pi$ iff $\mu^d$ is an eigenvalue of $\Pi^d, d \in \mathbb{N}$. From Lemma~\ref{lemNSA}:
\begin{itemize}
 \item[C1] $\exists \ell \in \mathbb{N}$ s.t. $\Pi^\ell$ has no zero entry in column corresponding to the absorbing state. Let $d_\star$ be the smallest such integer.
\item[C2] Every non-absorbing state has at least one zero element in the corresponding column of $\Pi^{d_\star}$.
\item[C3] Statements C1,C2  are true for any integer $d \geqq d_\star$.
\end{itemize}
We denote the column of ones as $\ones$, $i.e.$, $\ones = [1 \cdots 1]^T$ Since $\Pi^{d}$ is (row) stochastic, we have 
 $\Pi^{d} \ones =  \ones$. Hence, if $v$ is a left eigenvector for $\Pi^{d}$ with eigenvalue $\mu^{d}$, then:
\begin{gather}
v \Pi^{d} \ones = v\ones = \mu^{d} v \ones \Rightarrow (1-\mu^{d}) v \ones = 0
\end{gather}
implying that if $\mu^{d} \neq 1$, then $v\ones = 0$. Now we construct
 $\mathds{C} = [ C_1 \cdots C_n ]$, where $C_j = \min_j \Pi^{d}_{ij}$ (minimum column element). Considering the matrix $M=\Pi^{d} - \ones \mathds{C}$, we note:
\begin{gather}
( v\Pi^{d} = \mu^{d} v ) \wedge ( \mu^{d} \neq 1 )\Rightarrow v M = \mu^{d} v
\end{gather}
Recalling that stationary probability vectors (Perron vectors) of stochastic matrices add up to unity, we have:
\begin{gather}
 ( v\Pi^{d} =  v ) \Rightarrow v M = v - v\ones \mathds{C} = v - \mathds{C} 
\end{gather}
which, along with the fact that since $\mathds{C}$ is not a column of all zeros, implies that  an upper bound on the magnitudes of the eigenvalues of $M$ provides an upper bound on 
the magnitude of non-unity eigenvalues for $\Pi^{d}$. 
Now, invoking the 
Gerschgorin Circle Theorem~\cite{G31,V04}, we get:
\begin{gather}
 \vert \mu^{d} \vert \leqq 1 - \sum_j C_j = 1 - C_a \Rightarrow \vert \mu \vert \leqq \left ( 1 - C_a \right )^{\frac{1}{d}}
\end{gather}
where $C_a$ is the minimum column element corresponding to the  absorbing state. $1 - C_a$ is the maximum probability of not reaching the absorbing state after $d$ steps from any state, which is bounded above by
% :
% \begin{gather}
$ (a)^{d_1} ( b)^{d -d_1}$
% \end{gather}
 where $a$ is the maximum non-diagonal entry in $\Pi$ not going to the absorbing state, $b$ is the maximum of the non-unity diagonal entries in $\Pi$, and $d_1$ is a bounded integer. Since any sequence of non-selfloops is absorbed in a finite number of steps (strongly absorbing property), we have a finite bound for $d_1$. Hence we have:
\begin{gather}
 \vert \mu \vert \leqq \lim_{d \rightarrow \infty} a^{\frac{d_1}{d}} b^{1 - \frac{d_1}{d}} = b = \max_{q_i : \Pi_{ii} < 1} \Pi_{ii}
\end{gather}
This completes the proof.
\end{IEEEproof}
Next, we make rigorous our notion of policy performance, and near-global or $\epsilon$-optimality.
\begin{defn}[Policy Performance \& $\epsilon$-Optimality]\label{defperf} The performance vector $\rho^S$ of a given routing policy $S$  is the vector of node-specific probabilities of a packet eventually reaching the sink.
 A policy $U$ has \textit{Utopian performance} if its performance vector (denoted as $\rho^U$) element-wise dominates the one for  any arbitrary policy $S$, $i.e.$
% \begin{gather}
$ \forall q_i \in Q^N ,\rho^U_i \geqq \rho^S_i$.
% \end{gather}
A policy $P$ has \textit{$\epsilon$-optimal} performance, if for some given  $\epsilon > 0$, we have:
\begin{gather}
 \vert \vert \rho^P - \rho^U \vert \vert_\infty \leqq \epsilon
\end{gather}
% % 
\end{defn}

For a chosen $\theta$, the limiting policy $P_\theta$ computed by Algorithm~\ref{AlgorithmOPT} results in element-wise maximization of the measure vector over all possible
supervision policies (where supervision is to be understood in the sense of the defined control philosophy). $\widehat{\nu}^\infty_\theta$ 
is related to the policy performance vector $\rho^{P_\theta}$ as follows. Selective disabling of the transitions dictated by the policy $P_\theta$ induces a controlled PFSA, which represents the optimally supervised network, for a given $\theta$. Let the transition matrix for this optimized PFSA be $\Pi^\star_\theta$, and its Cesaro limit be $\Q^\star_\theta$. (Note:  $\Pi^\star_\theta$, $ \Q^\star_\theta$ are stochastic matrices.) Then:
\begin{gather}\label{eqperf}
 \forall q_i \in Q^N, \Q^\star_\theta \chi\big \vert_{i,(q_\Sn)} = \rho^{P_\theta}_i
\end{gather}
%
% We note here, that Eq.~\eqref{eqperf} implies that executing the optimization with vanishingly small $\theta$ will yield 
In the sequel, we would need to distinguish between the optimal measure vector  $\widehat{\nu}^\infty_{\theta'} $ (optimal for a given $\theta=\theta'$) computed by Algorithm~\ref{AlgorithmOPT}, and the one obtained by 
first computing $\widehat{\nu}^\infty_{\theta'} $ and then using the  PFSA  structure obtained in the process to compute the measure vector for some other value of $\theta=\theta''$.  These  two vectors may not be identical. 
\begin{notn}\label{not6}
In the sequel, we denote the vector obtained in the latter case as $\mudble{\theta'}{\theta''} $ implying that we have $ \mudble{\theta}{\theta} = \widehat{\nu}^\infty_{\theta}$.
\end{notn}
\begin{lem}\label{lem3} We have the following equalities:
\begin{subequations}
\begin{gather}
%  \forall \theta \in (0,1], \ \mudble{\theta}{\theta} = \widehat{\nu}^\infty_{\theta}\label{eqlemst1}\\
\lim_{\theta \rightarrow 0^+} \mudble{\theta'}{\theta} = \rho^{P_{\theta'}}\label{eqlemst2}\\
\lim_{\theta \rightarrow 0^+} \mudble{\theta}{\theta} = \rho^{U}\label{eqlemst3}
\end{gather}
\end{subequations}
\end{lem}
\begin{IEEEproof}
%  Eq.~\eqref{eqlemst1} is immediate. 
Recalling Eq.~\eqref{eqperf}, and noting that  for any PFSA with transition matrix $\Pi$ (with Cesaro limit $\Q$), we have $\lim_{\theta \rightarrow 0^+} \widehat{\nu}_\theta = \lim_{\theta \rightarrow 0^+}\theta \big [ \mathbb{I} - (1-\theta)\Pi \big ]^{-1} \chi = \Q \chi$, we have Eq.~\eqref{eqlemst2}.
In general, different choices of $\theta$ result in different disabling decisions, and hence different policies. However, since there is at most  a finite number of distinct policies for 
a finite network, there must exist a $\theta_\star$ such that for all choices $0 < \theta \leqq \theta_\star$, the policy remains unaltered (although the measure values may differ). Since, executing the optimization with vanishingly small $\theta$  yields a performance vector identical (in the limit) with the optimal measure vector  element-wise dominating the 
one for any arbitrary policy, the policy obtained for $0<\theta\leqq\theta_\star$ has Utopian performance. Hence:
\begin{gather}
 \lim_{\theta \rightarrow 0^+} \mudble{\theta}{\theta} = \lim_{\theta \rightarrow 0^+} \mudble{\theta_\star}{\theta} = \rho^{P_{\theta_\star}} = \rho^{U}
\end{gather}
This completes the proof.
\end{IEEEproof}
% 
% In general, different choices of $\theta$ result in different disabling decisions, and hence different policies. However, since there is at most  a finite number of distinct policies for 
% a finite network, there must exist a $\theta_\star$ such that for all choices $\theta \leqq \theta_\star$, the policy remains unaltered (although the measure values may differ).
% Since $\lim_{\theta \rightarrow 0^+} \widehat{\nu}_\theta = \lim_{\theta \rightarrow 0^+}\theta \big [ \mathbb{I} - (1-\theta)\Pi \big ]^{-1} \chi = \Q \chi$ for any PFSA with transition matrix $\Pi$ (with Cesaro limit $\Q$), it follows that as $\theta \rightarrow 0^+$, we have $\widehat{\nu}_\theta^\infty \rightarrow \rho^{P_\theta}$, which in turn implies:
% \begin{itemize}
%  \item[C4:] For all choices $\theta \leqq \theta_\star$, the limiting policy $P_\theta$ computed by Algorithm~\ref{AlgorithmOPT} has utopian performance in the sense of Definition~\ref{defperf}. 
% \end{itemize}
Computation of the critical $\theta_\star$ is non-trivial from a distributed perspective, although centralized approaches have been reported~\cite{CR07}. Thus it is hard to guarantee Utopian performance in Algorithm~\ref{AlgorithmOPT}. Also, $\theta_\star$ may be too small resulting in an unacceptably poor convergence rate.
 Nevertheless, we will show that, given any $\epsilon > 0$, one can choose $\theta$ to guarantee $\epsilon$-optimal performance of the limiting policy in the sense of Definition~\ref{defperf}. We would  need the following  result.
\begin{lem}\label{lemtechnical}
 Given any PFSA, with transition matrix $\Pi$ and corresponding Cesaro limit $\Q$, and  $\mu$ being a non-unity eigenvalue of $\Pi$  with  maximal magnitude, we have:
\begin{subequations}
% \vspace{0pt}
\calign[1pt]{
&\big \vert \big \vert \theta \big [ \mathbb{I} - (1-\theta)\Pi \big ]^{-1}- \Q \big \vert\big\vert_\infty \leqq \frac{\theta}{1-\vert \mu\vert }\label{eqclaim31}\\
% &\big \vert \big \vert   \theta\big [ \mathbb{I} - (1-\theta)\Pi \big ]^{-1}- \Q \big \vert\big\vert_1 \leqq \frac{\theta}{1-\vert \mu\vert }\label{eqclaim32}\\
 &\big\vert\big \vert\nu_{(\theta,\theta) } - \lim_{\theta' \rightarrow 0^+}\nu_{(\theta,\theta')}  \big\vert\big\vert_\infty \leqq \frac{\theta\vert \vert \chi \vert \vert_\infty}{1-\vert \mu\vert } \label{eqclaim33}
}
\end{subequations}
% 
% \vspace{2pt}
\end{lem}
\begin{IEEEproof}
Denoting $M=\big [ \mathbb{I} - (1-\theta)\Pi \big ]^{-1}- \frac{1}{\theta}\Q$, 
% \vspace{-10pt}
\begin{align}\textstyle
%  & \ \vert \vert\nu_\theta  - \Q\chi \vert\vert_1 \leqq \theta \big \vert \big \vert [\mathbb{I} - (1-\theta)\Pi ]^{-1}  - \Q \sum_{k=0}^\infty (1-\theta)^k\big\vert\big\vert_1 \vert \vert \chi \vert \vert_1\notag\\
 M = & [\mathbb{I} - (1-\theta)\Pi ]^{-1}  - \Q \sum_{k=0}^\infty (1-\theta)^k\notag\\\textstyle
 = &\sum_{k=0}^\infty (1-\theta)^k (\Pi -\Q)^k -\Q  
\notag\\
% &
= & [\mathbb{I} - (1-\theta)(\Pi - \Q)]^{-1} -\Q    \notag
\end{align}
We note, that if $u$ is a left eigenvector of $\Pi$ with unity eigenvalue, then $u\Q = u$. Also, if the eigenvalue corresponding to $u$ is strictly within the 
unit circle, then $u\Q = 0$. After a little algebra, it follows that  if $u$ is the left eigenspace (denoted as $E(1)$) corresponding to unity eigenvalues of $\Pi$, then 
$uM=0$, otherwise, $uM=\frac{1}{1-(1-\theta)\mu} u$, where $\mu$ is a non-unity eigenvalue for $\Pi$. Invoking the definition of induced matrix norms, and noting $\vert\vert A\vert \vert_\infty =\vert\vert A^T\vert \vert_1$ for any square matrix $A$:
\begin{gather}
 \vert \vert M \vert \vert_\infty = \max_{\vert \vert u \vert \vert_1 = 1} \vert \vert uM \vert \vert_1 = \max_{\vert \vert u \vert \vert_1 = 1 \wedge u \notin E(1)} \vert \vert uM \vert \vert_1
\end{gather}
We further note that since $[\mathbb{I} - (1-\theta)(\Pi - \Q)]^{-1}$ is guaranteed to be invertible~\cite{CR07}, its eigenvectors form a basis, implying:
% that we can represent any suitable $u$ as 
\begin{gather}\textstyle
 u = \sum_j c_j u^j, \ \textrm{with} \ \left \vert \left\vert \sum_j c_j u^j\right \vert\right \vert_1 = 1
\end{gather}
where $u^j$ are eigenvectors of $[\mathbb{I} - (1-\theta)(\Pi - \Q)]^{-1}$ with non-unity eigenvalues, and $c_j $ are complex coefficients.
 An upper bound for $\vert \vert M \vert \vert_1$ can be now computed as:
\begin{gather*}\textstyle
 \vert \vert M \vert \vert_\infty \leqq \frac{1}{1-(1-\theta)\vert \mu\vert } \left \vert \left\vert \sum_j c_j u^j\right \vert\right \vert_1 = \frac{1}{1-(1-\theta)\vert \mu\vert } \leqq  \frac{1}{1-\vert \mu\vert}
\end{gather*}
where $\mu$ is a  non-unity eigenvalue for $\Pi$ with maximal magnitude. This establishes Eq.~\eqref{eqclaim31}. 
% Considering right eigenvectors instead of left ones, and application of  essentially the same argument leads to Eq.~\eqref{eqclaim32}. 
Finally, noting:
\begin{gather*}
 \nu_{(\theta,\theta) } - \lim_{\theta' \rightarrow 0^+}\nu_{(\theta,\theta')} = \big (\theta[\mathbb{I} - (1-\theta)\Pi ]^{-1}  - \Q \big )\chi
\end{gather*}
establishes Eq.~\eqref{eqclaim33}.
% 
% 
% 
% Using this bound in Eq.~\eqref{eqBNd}, and noting that $\vert\vert \cdot \vert \vert_\infty \leqq  \vert\vert \cdot \vert \vert_1$, establishes the claim in Eq.~\eqref{eqclaim3}.
% % 
\end{IEEEproof}
The next proposition the  key result relating a specific choice of  $\theta$ to guaranteed $\epsilon$-optimal performance.
\begin{prop}[Global $\epsilon$-Optimality]\label{propglobal}
 Given any $\epsilon > 0$, choosing 
% $\theta$ as
% \begin{gather*}
$ \theta = \left.\epsilon\right/m^2 \ \textrm{where} \ m=\max_{q\in Q}\Crd(\mathcal{N}(q))$
% \end{gather*}
guarantees that the limiting policy computed by Algorithm~\ref{AlgorithmOPT} is  $\epsilon$-optimal in the sense of Definition~\ref{defperf}.
\end{prop}
\begin{IEEEproof}
We observe that the limiting measure values $\nuinf{ \vert_i} =  \nu^\star_\theta \vert_i$ computed by Algorithm~\ref{AlgorithmOPT} can be represented by convergent sums of the form ($a_{ij}$ are non-negative reals):
\begin{gather}\label{eqnondecreasing}
\forall q_i \in Q^N, \ \nuinf{ \vert_i} = \sum_{j=1}^\infty a_{ij} (1-\theta)^{j}
\end{gather}
implying that for each $q_i\in Q$, $\mudble{\theta}{\theta_1} \vert_i$ (See Notation~\ref{not6}) is a monotonically decreasing function of $\theta_1$ in the domain $[0,\theta]$.
We note that if the following statement:
\begin{gather*}
\forall q_i,q_j \in Q^N, \\  \nuinf{\vert_i} > \nuinf{\vert_j} \Rightarrow \forall \theta_1 \leqq \theta, \ \mudble{\theta}{\theta_1} \vert_i > \mudble{\theta}{\theta_1} \vert_j
\end{gather*}
is true, then we have Utopian performance for  policy $P_\theta$, $i.e.$, $\rho^{P_\theta} = \rho^U$. Hence, if $\rho^{P_\theta} \neq \rho^U$, then we must have:
\begin{gather*}
 \exists \theta_2 < \theta, \exists q_i,q_j \in Q^N, \\\big ( \nuinf{\vert_i} > \nuinf{\vert_j} \big )\wedge   \big ( \mudble{\theta}{\theta_1} \vert_i > \mudble{\theta}{\theta_1} \vert_j\big )
\end{gather*}
upon which Eq.~\eqref{eqnondecreasing}, along with the bound established in Eq.~\eqref{eqclaim31}, guarantees that if $q_i,q_j$ are nodes (in consecutive order) that satisfy the above statement, then:
\begin{gather}\label{eqbnd11}
 \lim_{\theta_1 \rightarrow 0^+} \left( \mudble{\theta}{\theta_1} \vert_i - \mudble{\theta}{\theta_1} \vert_j \right)\leqq \beta_\theta \theta
\end{gather}
where $\beta_\theta = \frac{1}{1 - \vert \mu \vert }$, with $\mu$ being a maximal non-unity eigenvalue of the transition matrix of the PFSA computed by Algorithm~\ref{AlgorithmOPT} at $\theta$.
Next we claim:
\begin{gather}
\forall \theta' \in (0, \theta],  \ \vert\vert \nuinf[\theta']{} - \mudble{\theta}{\theta'} \vert\vert_\infty \leqq m^2\theta
\end{gather}
We observe that, for any $\theta'$, the optimal policy $P_{\theta'}$ can be obtained by beginning with the PFSA induced  by $P_\theta$ (which is the optimal policy at $\theta$), and then executing the 
centralized iterative approach~\cite{CR07}, resulting in a sequence of \textit{element-wise non-decreasing} measure vectors converging to the 
optimal $\nuinf[\theta']{}$:
\begin{gather}
 \mudble{\theta}{\theta'} = \nu^{[0]}_{\theta'} > \nu^{[1]}_{\theta'}> \nu^{[2]}_{\theta'}>\cdots \nu^{[k^\star]}_{\theta'}=\nuinf[\theta']{}
\end{gather}
where $\nu^{[k]}_{\theta'}$ is the  vector obtained after the $k^{th}$ iteration, and $k^\star < \infty$ is the number of required iterations. 
Since, $
 \nu^{[k]}_{\theta'} = \theta'\big [ \mathbb{I} - (1-\theta')\Pi^{[k]}\big]^{-1}\chi
$, where  the transition matrix after $k^{th}$ iteration is $\Pi^{[k]}$ and setting $\Delta^{[k]}_{\theta'}= \nu^{[k]}_{\theta'} - \mudble{\theta}{\theta'}$ we have:
\begingroup\setlength\belowdisplayskip{0pt}
\begin{align}
 &\Delta^{[k]}_{\theta'} =  (1-\theta')\big [ \mathbb{I} - (1-\theta')\Pi^{[k]}\big]^{-1} (\Pi^{[k]} - \Pi^{[0]}) \mudble{\theta}{\theta'} \notag \\ 
  &= {\textstyle\frac{1-\theta'}{\theta'}}\big \{ {\red \underbrace{\black \theta' \big [ \mathbb{I} - (1-\theta')\Pi^{[k]}\big]^{-1} }_{\red \mathds{B}^{[k]}_{\theta'}}}\big \}\big \{ {\red \underbrace{\black(\Pi^{[k]} - \Pi^{[0]}) \mudble{\theta}{\theta'}}_{\red \omega^{[k]}_{\theta'}}}\big \}\notag
\vspace{-15pt} 
\end{align}
\endgroup
%
%
% \the\belowdisplayskip \\
For $q_i \in Q$, let ${\upd}_i^{(0\rightarrow k)}$ be the set of transitions $(q_i \xrightarrow{\sigma} q_j)$, which are updated ($i.e.$ enabled if disabled or vice verse) to go from the configuration corresponding to $\nu^{[0]}_{\theta'}$ to the one corresponding to $\nu^{[k]}_{\theta'}$.
We note that:
\begin{gather*}
 \upd_i^{(0\rightarrow k)} = \left ( \upd_i^{(0\rightarrow 1)} \cap \upd_i^{(0\rightarrow k)} \right ) \bigcup \mathscr{W}
\end{gather*}
where $\mathscr{W} = \upd_i^{(0\rightarrow k)} \setminus \left ( \upd_i^{(0\rightarrow 1)} \cap \upd_i^{(0\rightarrow k)} \right )$. 
The $i^{th}$ row of  $\Pi^{[1]}$ is obtained from $\Pi^{[0]}$~\cite{CR07} by disabling controllable transitions $q_i \xrightarrow{\sigma} q_j$ if $\nu^{[0]}_{\theta'} \vert_j > \nu^{[0]}_{\theta'} \vert_i$ (and enabling otherwise), and each such update leads to a positive contribution in the corresponding row of $\omega^{[1]}_{\theta'}$. It follows that 
updating any transition $t \equiv (q_i \xrightarrow{\sigma} q_j) \in \left ( \upd_i^{(0\rightarrow 1)} \cap \upd_i^{(0\rightarrow k)} \right )$ leads to a positive contribution to $\omega^{[k]}_{\theta'}\vert_i$, given by:
\begin{gather}
 C_t = \Pitilde (q_i ,\sigma) \left \vert \nu^{[0]}_{\theta'}\big \vert_i -\nu^{[0]}_{\theta'}\big \vert_j \right \vert  
\end{gather}
and for every transition $t' \equiv (q_i \xrightarrow{\sigma'} q_k) \in \mathscr{W}$ leads to a negative contribution to $\omega^{[k]}_{\theta'}\vert_i$, given by:
\cgather{
 C_{t'} = -\Pitilde (q_i ,\sigma') \left \vert \nu^{[0]}_{\theta'}\big \vert_i -\nu^{[0]}_{\theta'}\big \vert_k \right \vert  \\
\mspace{-90mu} \textrm{implying that:} \mspace{13mu} 
 \omega^{[k]}_{\theta'}\vert_i \leqq \mspace{-140mu} \sum_{\mspace{180mu} r \in \left ( \upd_i^{(0\rightarrow 1)} \bigcap \upd_i^{(0\rightarrow k)}\right ) } \mspace{-140mu} C_r  \mspace{92mu} \\
\Rightarrow \omega^{[k]}_{\theta'}\vert_i \leqq \sum_{\sigma \in \Sigma} \Pitilde(q_i,\sigma) \beta_\theta \theta  = \beta_\theta \theta  \mspace{20mu} \textrm{(See Eq.~\eqref{eqbnd11})}\notag
}
Since the rows corresponding to the absorbing states have no controllable transitions, absorbing states must remain absorbing through out the iterative sequence, and the corresponding entries in $\omega^{[k]}_{\theta'}$ for all $k\in\{0,\cdots,k^\star\}$ are strictly $0$. It follows:
\begin{align}
 &\omega^{[k]}_{\theta'}\vert_i = \left \{ \begin{array}{ll}
0 & ,\textrm{if $q_i$ is absorbing}\\
       \in [0,  \beta_\theta\theta] & ,\textrm{otherwise }
                              \end{array}
\right.\label{eq52}
\end{align}
Stochasticity of $\mathds{B}^{[k]}_{\theta'}$ implies that in the limit $\theta' \rightarrow 0^+$, $\mathds{B}^{[k]}_{\theta'}$ converges to the 
Cesaro limit of $\mathds{B}^{[k]}_{\theta'}$. Applying Lemma~\ref{lemtechnical}:
\begin{gather}
\big \vert \big \vert \mathds{B}^{[k]}_{\theta'} - \lim_{\theta' \rightarrow 0^+}\mathds{B}^{[k]}_{\theta'}\big \vert \big \vert_\infty \leqq \frac{\theta'}{1-\vert \mu_{\theta'}\vert } \triangleq \beta_{\theta'}\theta'
\end{gather}
where $\mu_{\theta'}$ is a non-unity eigenvalue for $\mathds{B}^{[k]}_{\theta'}$ with maximal magnitude. Using the invariance of the absorbing state set,
and observing that the Cesaro limit $\lim_{\theta' \rightarrow 0^+}\mathds{B}^{[k]}_{\theta'}$ has strictly zero columns corresponding to non-absorbing states, we conclude:
\begin{gather*}
 \forall \theta' \in(0,\theta], \ \Delta^{[k]}_{\theta'}\vert_i \leqq \frac{1-\theta'}{\theta'}\beta_{\theta'}\theta'\beta_{\theta}\theta \leqq \beta_{\theta'}\beta_{\theta}\theta
\end{gather*}
It is easy to see that the PFSA induced by $P_\theta$ is strongly absorbing (Definition~\ref{defNSA}), and so is each one obtained in the iteration. 
Also, the virtual nodes in our network model have no controllable transitions, and have no self-loops. Physical nodes can have self-loops arising from disablings;
but for a non-absorbing node with at most $m$ neighbors, the self-loop probability is bounded by $(m-1)/m$, which then implies $\beta_{\theta'},\beta_\theta \leqq \frac{1}{1 - (m-1)/m} = m$ (Proposition~\ref{propspectral}).
Hence:
\begin{gather}
 \forall \theta' \in (0,\theta], \ \vert \vert \Delta^{[k]}_{\theta'}\vert\vert_\infty \leqq m^2 \theta
\end{gather}
Thus, if we choose $\theta = \epsilon / m^2$, we can argue:
\begin{align*}
  &\forall k \in \{0,\cdots,k^\star\}, \  \forall \theta' \in (0, \theta], \ \vert \vert\Delta^{[k]}_{\theta'} \vert\vert_\infty \leqq  \epsilon  \\
 \Rightarrow & \lim_{\theta' \rightarrow 0^+} \left \vert \left \vert \widehat{\nu}^{\infty}_{\theta'} - \mudble{\theta}{\theta'} \right \vert \right \vert_\infty \leqq \epsilon \\ 
\Rightarrow  & \left \vert \left \vert \lim_{\theta' \rightarrow 0^+}\widehat{\nu}^{\infty}_{\theta'} - \lim_{\theta' \rightarrow 0^+}\mudble{\theta}{\theta'} \right \vert \right \vert_\infty \leqq \epsilon  \ \left ( \textrm{\small \sffamily \txt{{Continuity}\\ {of norm}}} \right )\notag \\ 
 \Rightarrow & \left \vert \left \vert \rho^U - \rho^{P_\theta} \right \vert \right \vert_\infty \leqq \epsilon \ \left ( \textrm{\small \sffamily {Using Lemma~\ref{lem3}}} \right )\notag
\end{align*}
which completes the proof.
\end{IEEEproof}
Once we have guaranteed convergence to a $\epsilon$-optimal policy, we need to compute asymptotic bounds on the time-complexity of route convergence, $i.e.$, how long it takes to 
converge to the limiting policy so that the local routing decisions no longer fluctuate. In practice, the convergence time is dependent on the network delays, the degree to which the node updates are synchronized $etc.$, and is difficult to estimate. In this paper, we neglect such effects to obtain an asymptotic estimate in the perfect situation. This allows us to quantify the  dependence of the convergence time on key parameters such as $N$, $m$ and $\epsilon$. Future work will address situations where such possibly implementation-dependent effects are explicitly considered resulting in  potentially smaller convergence rates.
\begin{prop}[Asymptotic Runtime Complexity]\label{propcomplex}With no communication delays and assuming synchronized updates,  convergence time $\Tc $ to $\epsilon$-optimal operation for a network of $N$ physical nodes and maximum $m$ neighbors, satisfies:
\begin{gather*}
  \Tc  = O\left (  \frac{Nm^2}{\epsilon (1-\gamma_\star)}\right ) \\ \textrm{where $\gamma_\star$ is a lower bound on drop probabilities} 
 \end{gather*}
\end{prop}
\begin{IEEEproof}
 Synchronized updates imply that we can assume  the measure vector to update via the following recursion:
\begin{subequations}
\begin{align}
 \widehat{\nu}^{[1]}_\theta  &= \boldsymbol{0} \ \textrm{(Zero vector)}\\
 \widehat{\nu}^{[k+1]}_\theta  &= (1-\theta)\Pi^{[k]} \widehat{\nu}^{[k]}_\theta + \theta\chi
\end{align}
\end{subequations}
which can be used to obtain the upper bound:
\begin{gather}
\big \vert \big \vert \widehat{\nu}^{\infty}_\theta - \widehat{\nu}^{[k]}_\theta  \big \vert \big \vert_\infty \leqq (1-\theta)^k
\end{gather}
implying that after $k$ updates, each node is within $(1-\theta)^k$ of its limiting value. Denoting the smallest difference of measures as $\Delta_\star$, we note that 
$(1-\theta)^k\leqq \Delta_\star$ would guarantee that no further route fluctuation occurs, and the network operation will be $\epsilon$-optimal from that point onwards. To estimate $\Delta_\star$, we note that 1) comparisons cannot be made for values closer than the machine precision $M_0$, and 2) 
the lowest possible non-zero measure in the network occurs at the network boundaries if we assume the worst case scenario in which the drop probability is always $\gamma_\star$. We recall  
the measure of a node is  the sum of the measures of all paths initiating from the particular node and terminating at the sink. Also, note that 
any such path accumulates a multiplicative factor of $(1-\theta)^2(1-\gamma_\star)$ in each hop. Assuming the worst case, where a given node is $N$ hops away, and has a single path to the sink, we conclude that the smallest non-zero measure of any node is bounded below by $( (1-\theta)^2(1-\gamma_\star))^N$, inducing the following bound:
\begin{gather}
 \Delta_\star \geqq M_0 \left( (1-\theta)^2(1-\gamma_\star)\right )^N
\end{gather}
and hence a sufficient condition for convergence is:
\begin{align}
 &(1-\theta)^k = M_0 \left( (1-\theta)^2(1-\gamma_\star)\right )^N \notag\\
\Rightarrow &(1-\theta)^{(k-2N)} = M_0 (1-\gamma_\star)^N \notag\\
\Rightarrow &k = 2N + \frac{\log M_0}{\log (1-\theta)} + N \frac{\log (1-\gamma_\star)}{\log (1-\theta)} 
\end{align}
Treating $M_0$ as a constant, we have 
%  \begin{gather}
 $ \frac{\log M_0}{\log (1-\theta)} = O\left (\frac{1}{\theta}\right )$.
%  \end{gather}
Since $\theta$ must be small for near-optimal operation and considering the worst case $\gamma_\star \ll 1$, we have: 
\begin{align}
&(1-\theta)^{k_1} = 1 - \gamma_\star \mspace{20mu} \textrm{where }k_1 \triangleq \frac{\log (1-\gamma_\star)}{\log (1-\theta)} \notag\\
\Rightarrow &(1-k_1\theta) \simeq 1 - \gamma_\star \Rightarrow  k_1 \theta = \gamma_\star 
\Rightarrow k_1 = \frac{\gamma_\star}{\theta} \notag\\ \Rightarrow & k_1 \simeq \frac{1}{\theta (1 - (1-\gamma_\star))^{-1}}\Rightarrow k_1  =O\left( \frac{1}{\theta( 1-\gamma_\star)}\right)\notag\\
% \begin{gather}
% \end{gather}
% 
% \begin{align}
\Rightarrow & k = O\left(N+ \frac{1}{\theta} +  \frac{N}{\theta(1-\gamma_\star)}\right ) = O\left( \frac{N}{\theta(1-\gamma_\star)}\right )\notag\\
\Rightarrow & k = O\left( \frac{Nm^2}{\epsilon(1-\gamma_\star)}\right ) \mspace{20mu} \mathsf{(Using \ Proposition~\ref{propglobal})} \notag
\end{align}
Under the assumption of no communication delay, we have $\Tc  = O(k)$, which completes the proof.
\end{IEEEproof}
It follows from Proposition~\ref{propcomplex} that for constant $\epsilon$ and $\gamma_\star$, and large networks with relatively smaller number of local neighbors such that $N \gg m$, we will have 
% \begin{gather*}
$ \Tc  =O(N)$. 
% \end{gather*}
{\itshape Detailed simulation, on the other hand, indicates that this bound is not tight, as illustrated in Figure~\ref{figcomplex}(a), where we see a logarithmic dependence instead.}
% Next, we discuss the key characteristics of  implementation specifics.
% #################################################
% #################################################
% % % % #################################################
%#############################################
\begin{table}[t]
% \normalsize
\caption{Instantaneous Node Data Table For GODDeS}\label{tableNodedata}
\vspace{-15pt}
\centering
\begin{minipage}{3.1in}
\footnotesize\parashade[0.97]{sharpcorners}{\vspace{-2.1pt}\begin{tabular}{p{.05in}|c|c|c|c}
\sffamily \txt{Id.}&\sffamily \txt{Neighbor  \#}& \sffamily \txt{Current \\ Measure} & \sffamily \txt{$\phantom{^1}$Drop \\ $\phantom{_1}$Probability} & \sffamily \txt{Forwarding \\ Decision }\\\hline
% \rowcolor[gray][0.1]
{$I_1$}& \BRed {{\scshape (Self)} $1$} & \bf \BRed {$\nu_0$} & \bf \BRed {$d_0=0$} & \bf \BRed {$0$}\\ \hline
$\vdots$&$\vdots$ & $\vdots$ & $\vdots$ & $\vdots$ \\ \hline
{$I_m$}&$m$ & $\nu_m$ & $d_m$ & $1$\\ %\hline
 \end{tabular}
\vspace{-2pt}}
\end{minipage}
\end{table}
%#############################################
\section{Properties \& Implementation Details}\label{sec5}
%#############################################
The GODDeS pseudo-code in Algorithm~\ref{AlgorithmOPT} specifies the  instructions executing  on each physical node,  in an asynchronous and distributed manner.
By design, GODDeS only uses information that is locally available, and global performance guarantees are achieved by propagating this local information via neighbor-neighbor communication. The idea of such information percolation in networks is not particularly new; the novelty of GODDeS lies in the exploitation of sound theoretical results from language measure theory to 
design such communication. The node-specific measure values computed by GODDeS essentially reflects a {\itshape generalized distance vector}, that takes in to account link-specific drop probabilities which update as network statistics ($e.g.$ the drop probabilities) change (albeit at a slower time scale). Using the notion of quantitative measures of probabilistic regular languages, GODDeS successfully integrates the well-known notions of distance vector and link states into one single node-specific scalar; namely the measure at each node. 
Thus  the amount of data that needs to be communicated is very small, implying a low communication overhead. Updating these measure values is also very simple, as stipulated in Algorithm~\ref{AlgorithmOPT}. 
Routing then proceeds by local multi-casting to neighbors which currently have a strictly higher measure; and our theoretical results guarantee that such a policy will essentially result in $\epsilon$-optimal global  performance. Furthermore, as we show in Proposition~\ref{propchar}, the optimal routing policy is inherently free from loops and the formidable count-to-infinity problem.
% We summarize these observations as:
%################################################################################################################
\begin{figure}[t]
\centering
% \psfrag{N}[cr]{\footnotesize \txt{N$\phantom{x}$}}
% \psfrag{T}[c][c][1][90]{\footnotesize $\phantom{XXX}$\txt{Time \\(ms)$\phantom{x}$}}
% \psfrag{2}[cc]{\footnotesize  \txt{$\phantom{x}$\\$10^2$}}
% \psfrag{3}[cc]{\footnotesize  \txt{$\phantom{x}$\\$10^3$}}
% \psfrag{200}[cr]{\footnotesize  \txt{200}}
% \psfrag{400}[cr]{\footnotesize  \txt{400}}
% \psfrag{600}[cr]{\footnotesize  \txt{600}}
% \psfrag{800}[cr]{\footnotesize  \txt{800}}
% \psfrag{1000}[cr]{\footnotesize  \txt{1000}}
% \psfrag{1200}[cr]{\footnotesize  \txt{1200}}
% \psfrag{1400}[cr]{\footnotesize  \txt{1400}}
% \psfrag{220}[cr]{\footnotesize  \txt{220}}
% \psfrag{Mean123456789}[cl]{\scriptsize \txt{Mean\\$\phantom{X}$}}
% \psfrag{x}[cl]{\scriptsize \txt{}}
%  \psfrag{B}[c][c][1][90]{\footnotesize $\phantom{XXX}$\txt{Time (ms)$\phantom{x}$}}
% \subfigure[]{\includegraphics[width=2.75in]{Figures/complexity}}
%  \psfrag{B}[c][c][1][90]{\footnotesize $\phantom{XXX}$\txt{Time\\ (ms)$\phantom{x}$}}
% \psfrag{2.3}[br]{\footnotesize \txt{\\$10^{3.3}$}}
% \psfrag{2.4}[cr]{\footnotesize \txt{\\$10^{3.4}$}}
% \psfrag{-5}[lc]{\footnotesize \txt{$\phantom{x}$\\$10^{-5}$}}
% \psfrag{-4}[lc]{\footnotesize \txt{$\phantom{x}$\\$10^{-4}$}}
% \psfrag{-3}[lc]{\footnotesize \txt{$\phantom{x}$\\$10^{-3}$}}
% \psfrag{N}[cr]{ \txt{$\epsilon\phantom{x}$}}
% \psfrag{MS}[lc]{ \small\txt{Smoothing \\Spline Fit}}
%  \subfigure[]{ \includegraphics[width=2.75in]{Figures/complexity1}}
\includegraphics[width=3.25in]{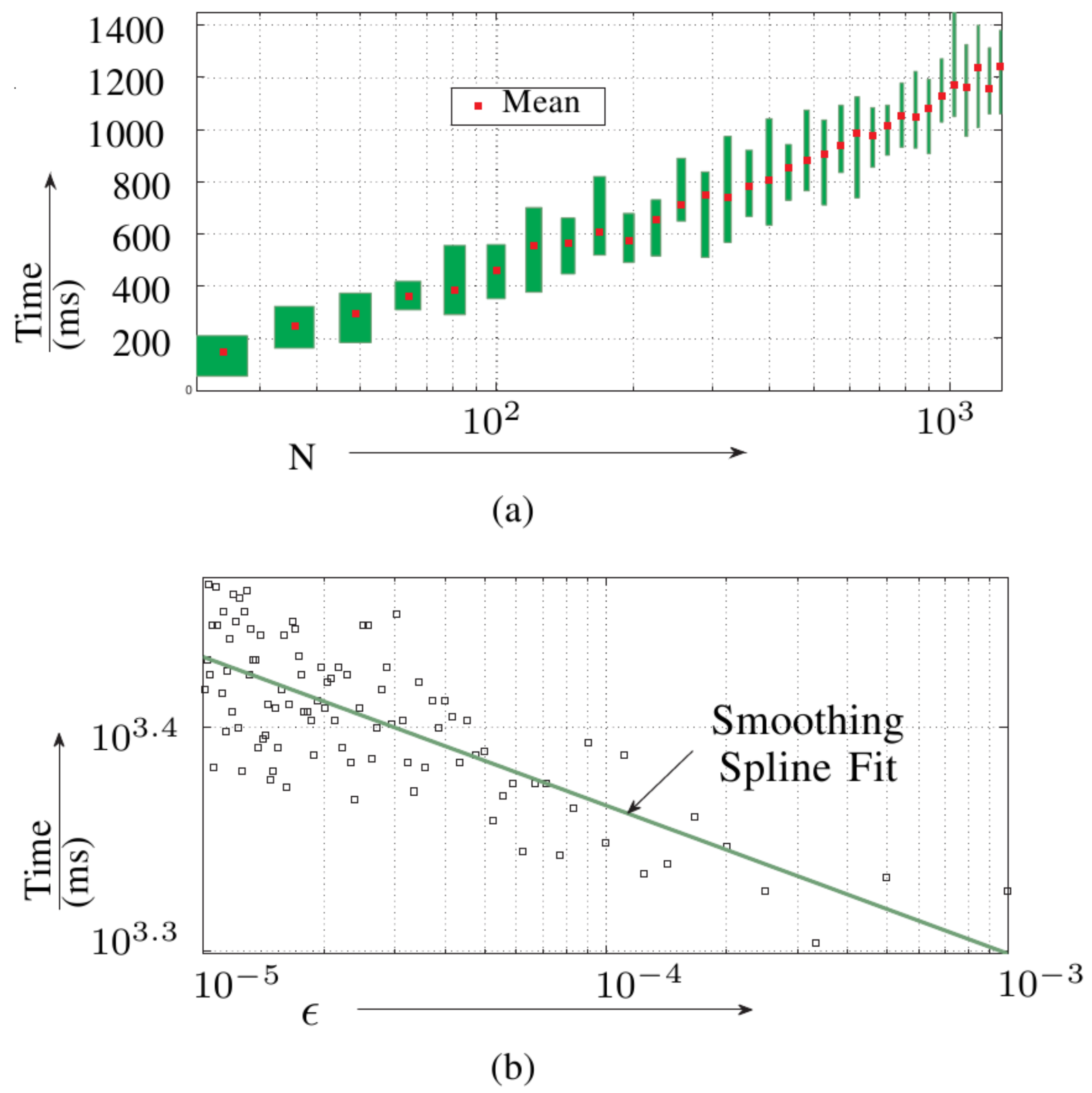}
\caption{Convergence complexity: (a) illustrates little dependence of convergence on network size. (b) captures the $O(1/\epsilon)$ dependence}\label{figcomplex}
\end{figure}
%#############################################
%#############################################
\begin{prop}[Properties]\label{propchar}
     The limiting GODDeS policy: 
     \begin{enumerate}
          \item is loop-free
 \item  is the unique loop-free policy that disables the smallest set of transitions among all policies which induce the same measure vector for a given $\theta$. 
\end{enumerate}
\end{prop}
\begin{IEEEproof}
(1) Absence of loops follows immediately from noting that, 
 in the limiting policy, a controllable transition $q_i \rightarrow q^v_{(ij)}$ is enabled if and only if $q^v_{(ij)}$ has a limiting measure strictly greater than that of  $q_i$, implying that any sequence of transitions (with no consecutive repeating states) goes to either the dump  or the sink in  a finite number of steps.

(2) follows directly from the uniqueness and the maximal permissivity property of optimal policies computed by language measure-theoretic optimization (See \cite{CR07}).
\end{IEEEproof}
% 
% No loops
% 
% Multi-path optimality
% 
% Uniqueness
% %#############################################
% \begin{figure}[t]
% \centering
% \psfrag{G}[cc]{\bf \sffamily \footnotesize \color{CadetBlue4}\txt{GODDeS \\ Updates\\ (Algo. 1)}}
% \psfrag{B}[cc]{\bf \sffamily \footnotesize \color{Green4}\txt{Information \\ Exchange\\ With \\ Neighbors}}
% \psfrag{S}[cc]{\bf \sffamily \scriptsize \color{IndianRed2}\txt{Asynchronous \\ Operation}}
% \psfrag{H}[cc]{\bf \sffamily \footnotesize \txt{Node Data\\ Table}}
% \psfrag{A}[cc]{\bf \sffamily \footnotesize \color{DodgerBlue2}\txt{Routing \\ Decision}}
% % \psfrag{N}[cc]{\bf \sffamily \footnotesize \color{DodgerBlue2}\txt{Intra-node Operational Scheme}}
%      \includegraphics[width=2.5in]{Figures/goddes1}
%      \caption{Intra-node Operational Scheme For GODDeS }\label{figgoddes}
% \end{figure}
% %#############################################

% %#############################################
%#############################################
\begin{figure}[t]
\centering
\includegraphics[width=3.25in]{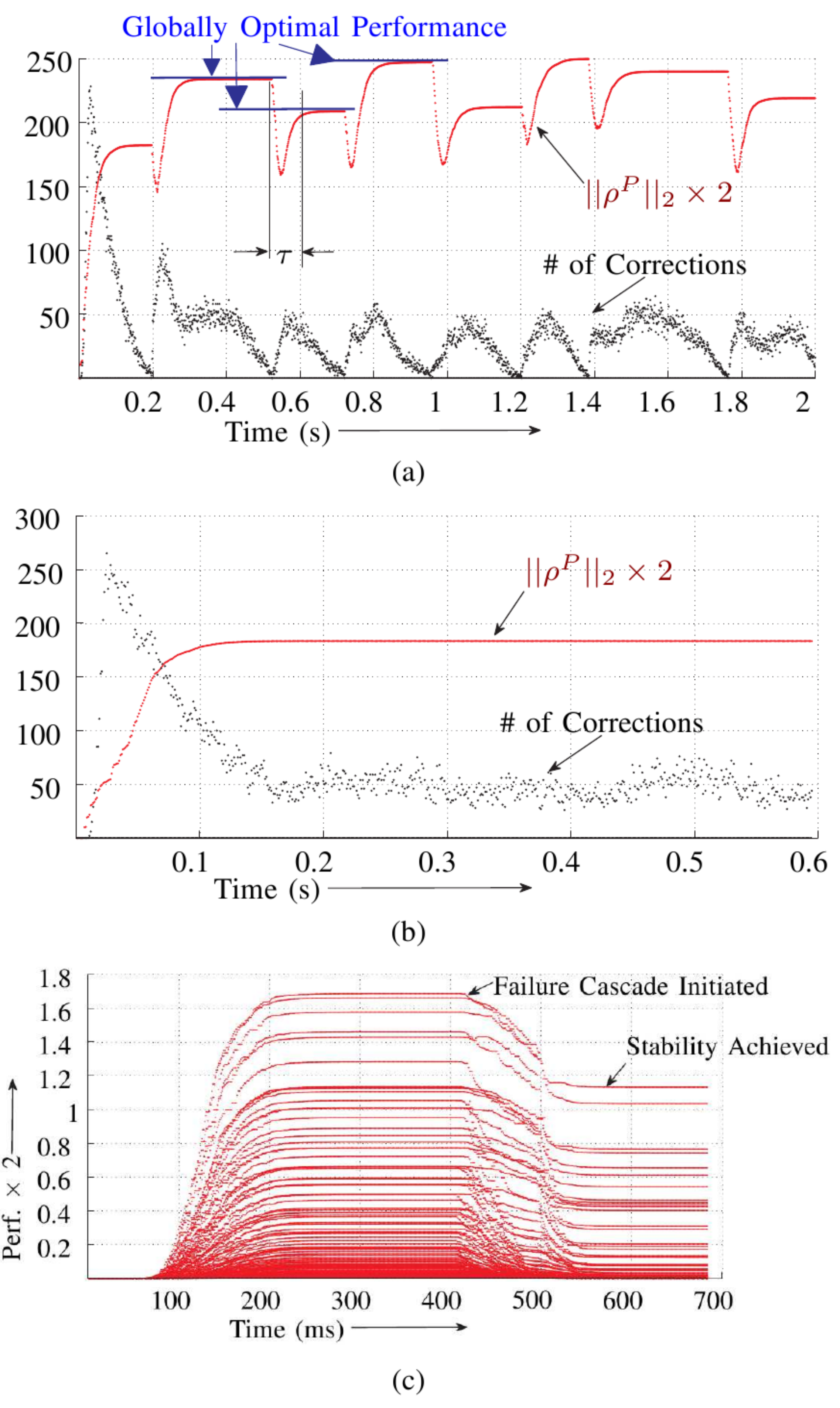}
% \end{minipage}
\caption{Convergence dynamics: (a) rapid convergence to large random sink movements  (b)robust response to  large zero-mean variations in the drop probabilities (c) response to failure cascade where $50\%$ of the nodes are killed}\label{figsim1}
\end{figure}
%#############################################
%#############################################
In this paper, we refrain from explicitly designing specific headers and data-structures that would be required for practical implementation of GODDeS. However one can  easily  tabulate the  data that needs to be maintained at each node  (See Table~\ref{tableNodedata}). In particular, each node needs to know the unique network id. of each neighbor that it can communicate with (Col. 1), and their current measure values (Col. 3). The drop probabilities for communicating from self to each of those neighbors must be maintained as well, for the purpose of carrying out the GODDeS updates (Col. 4). The forwarding decision is a neighbor-specific Boolean value (Col. 5), which is set to $1$ if the neighbor currently has a strictly higher measure than self, and $0$ otherwise. The packets are then forwarded by randomly choosing (in an equiprobable manner) between  the enabled neighbors, $i.e.$, the ones with a \textbf{true} forwarding decision.
Note that this node data  updates when the measures of the neighbors change (Col. 3), or the drop probabilities (Col. 4) update. However, changes in the measures may not necessarily reflect a change in the forwarding decisions. Also, note that the routing is inherently probabilistic, (due to the possibility that multiple enabled neighbors may exist for a given node). Furthermore, the optimal policy disables transmission to as few neighbors as possible for a specified $\theta$  (Proposition~\ref{propchar}),  and hence exploits multi-path transmissions in an optimal manner.
% 
%#############################################
%#############################################
\begin{figure*}[t]
\centering
% \psfrag{S}[cr]{\small \bf \txt{Sink}}
% \psfrag{D}[cc]{\small  \txt{Dead\\ Nodes}}
% \psfrag{P}[cb]{\color{white}\bf \txt{Poor}}
% \psfrag{T}[lc]{\color{Red4}\bf \txt{Time}}
\includegraphics[width=6.5in]{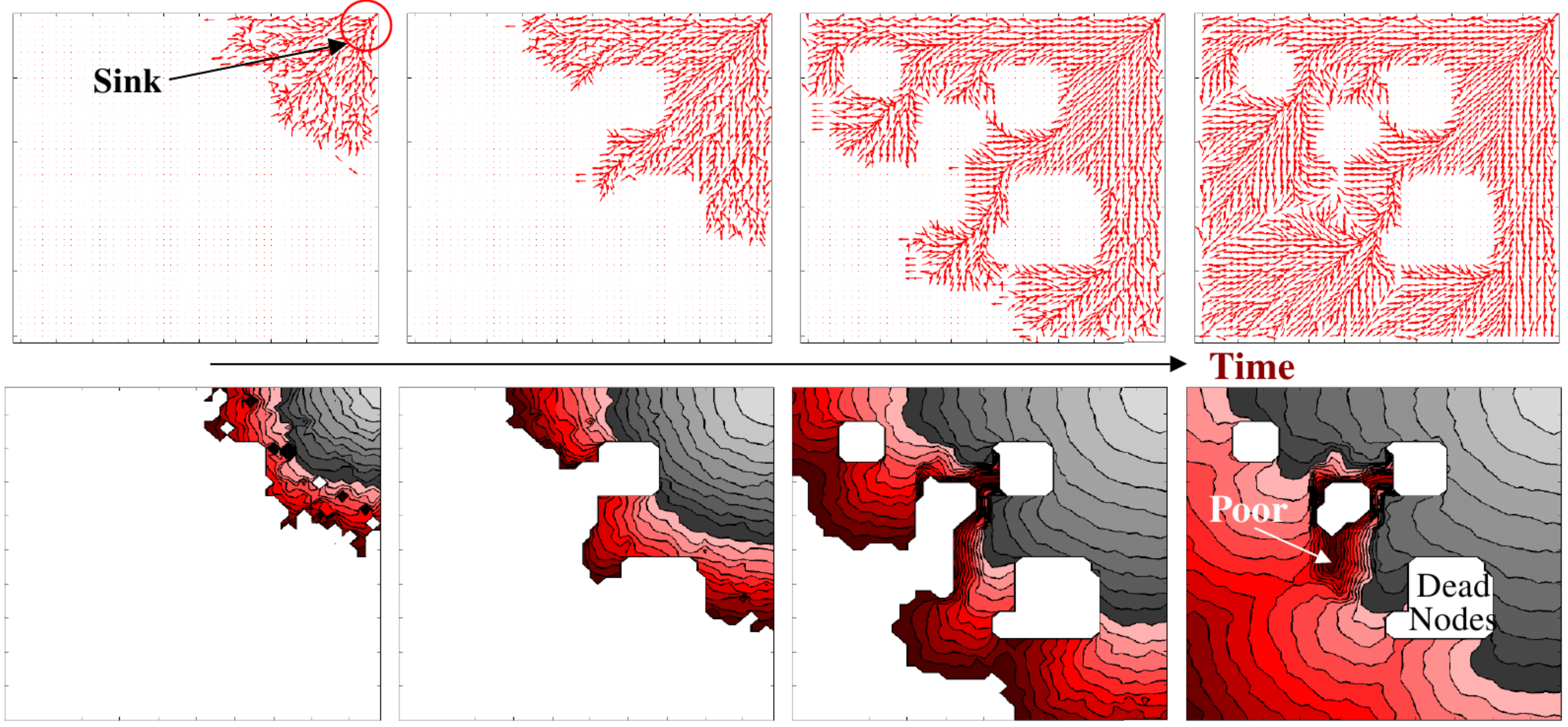}
\caption{Convergence dynamics: Top plates illustrate the gradient of the scalar field induced by the measure values; bottom plates illustrate the level sets}\label{figsim2}
\end{figure*}
%#############################################
%#############################################
%#############################################
% \vspace{-10pt}
% \subsection{Energy, Congestion, \& Other Optimization Objectives}

In remote sensing applications nodes often have  limited energy, necessitating route updates as high-traffic nodes get depleted. 
Also, local congestion arising due to the bursty nature of such communication may require re-routing. 
Note that congestion leads to higher packet drop probabilities, and gets reflected in the local link-specific drop probability estimations. Thus,  GODDeS automatically corrects 
for network congestion to a large degree, by modulating the forwarding decisions as specific areas experience high traffic. However this does not correct for depleting energy levels (until the nodes actually die). Energy-aware reorganizations can be nevertheless carried out  within the GODDeS framework autonomously and in a decentralized manner. 
Specifically, each node can regulate incoming traffic by deliberately reporting  lower values of its current self-measure to its neighbors: 
\begin{gather}\label{eqfactor}
\textrm{\small \sffamily \BRed Reported} \longrightarrow r^{[k]}_{\theta}\big \vert_i     = \zeta(q_i,k)\nu^{[k]}_{\theta}\big \vert_i \leftarrow \textrm{\small \sffamily \Dgreen Computed}
\end{gather}
where $\forall q_i \in Q, k \in[0,\infty), \zeta(q_i,k) \in [0,1]$ is a multiplicative factor which is modulated to have  decreasing values as node energy gets depleted, or as local congestion increases. Such modulation  forces automatic self-organization to compute alternate routes that tend to avoid the particular node. The dynamics of such context-aware modulation  may  be non-trivial; while for slowly varying $\zeta(q_i,k)$, the convergence results presented here is expected to hold true, rapid fluctuations in  $\zeta(q_i,k)$  may be problematic.
%################################################################################################################
\section{Verification, Validation \& Discussion}\label{sec6}
Extensive simulations have been performed on NS2 network simulator, running on a 32  core (64 bit architecture) workstation with 128 GB of RAM.   We investigate how  convergence times scale as a function of the network size in Figures~\ref{figcomplex}(a-b). $10^2$ random topologies were considered for each $N$ (increased from 25 to 1600), and the mean times along with the max-min bars are plotted in Figure~\ref{figcomplex}(a).
Note that the abscissa is on a logarithmic scale, and the near linear nature of the plot indicates a logarithmic dependence of the convergence on network size, implying that the bound computed in Proposition~\ref{propcomplex} is possibly not tight. The dependence on $\epsilon$ shown in Figure~\ref{figcomplex}(b) (for $N=10^3$) is hyperbolic, as expected, leading to a near linear dependence after a smoothing spline fit on a log-log scale. Note the convergence times are not CPU times, but are estimated from NS2 output (using  802.11 standard).
%#############################################
%#############################################
\begin{figure*}[t]
\centering
\includegraphics[width=6.5in]{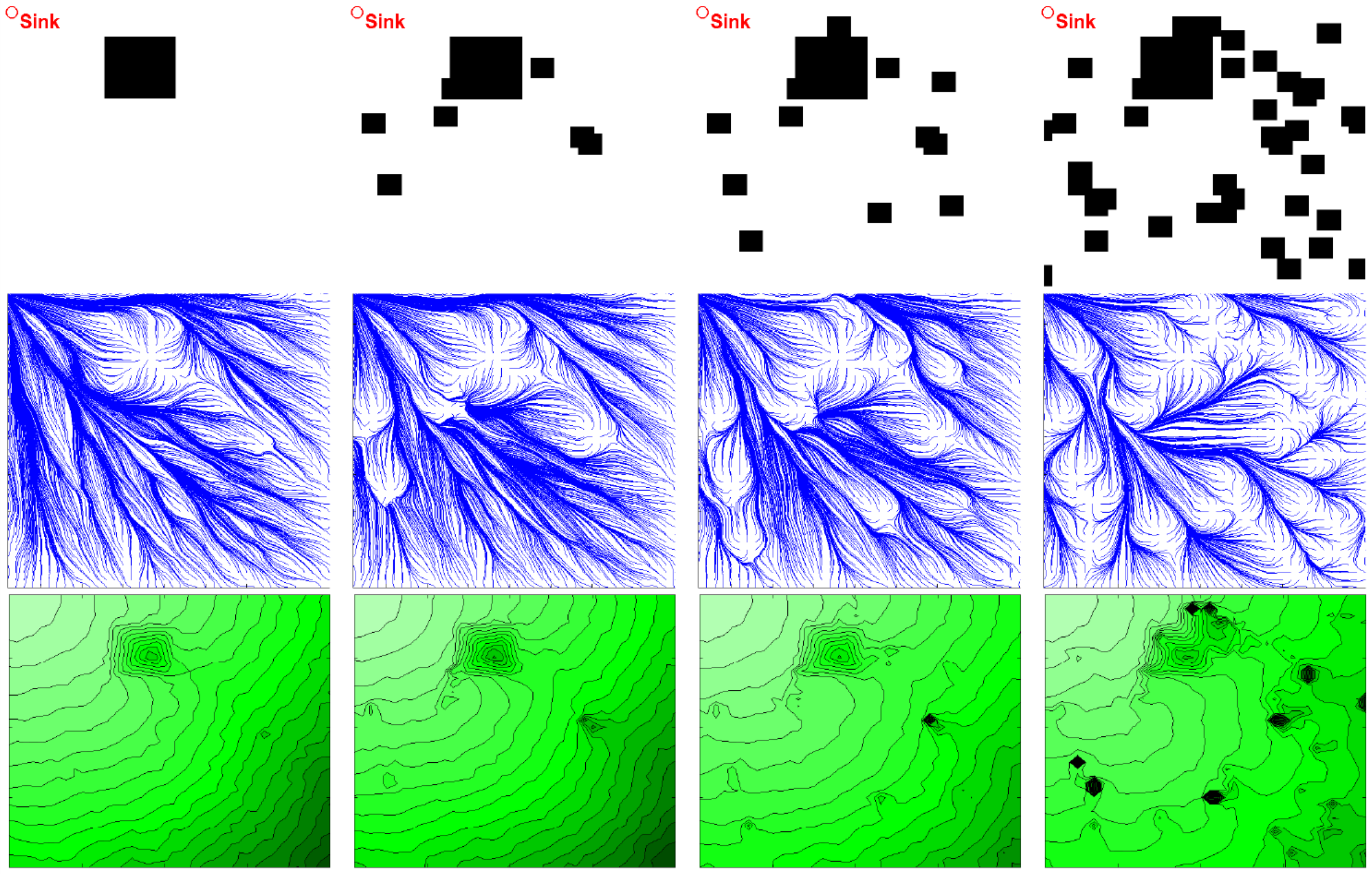}
\caption{Time lapse plates for progressive node deaths: Top row indicates failed regions in black, middle row illustrates packet path signatures to the sink from operational nodes, and bottom row shows the level sets for the scalar field  induced by the node measures}\label{figsim3}
\end{figure*}
%#############################################
%#############################################

The theoretical convergence results are illustrated in Figure~\ref{figsim1}(a-c), which were generated on a $10^4$ node network. Plate (a) illustrates the variation of the number of route updates ($\#$ of forwarding decision corrections) and the norm of the performance vector $\rho^P$ (scaled up by a multiplicative factor of 2) when the sink is moved around randomly at a slower time scale. Since $\rho^P$ is the vector of end-to-end success probabilities (See Definition~\ref{defperf}), its norm captures the degree of expected  throughput across the network. Note that sink changes induce self-organizing corrections, which rapidly die down, with the performance converging close to the global optimal ($\epsilon=0.001$ was assumed in all the simulations). The drop probabilities are chosen randomly, and, on the average, held  constant  in the course of simulation illustrated in plate (a) (zero mean Gaussian noise is added to illustrate robustness). Note that the seemingly large fluctuations in the performance norm is unavoidable; the interval $\tau$ is the what it approximately takes for information to percolate through the network, and hence  this much time is necessary at a minimum for decentralized route convergence. Plate (b) illustrates the effect of large zero-mean stochastic variations in the drop probabilities. Each node estimates the drop probabilities from simple windowed average of the link-specific packet drops. We note that large sustained  fluctuations result in a sustained corrections in the forwarding decisions (which no longer goes to zero). However, the norm of the performance vector converges and holds steady, indicating a highly stable quality of service.  This clearly illustrates that the information percolation strategy induces a low-pass filter eliminating high-frequency  fluctuations, yielding a self-organizing routes that maintain high throughput in a robust manner. Note that small number of route fluctuations always occur (as shown by the non-zero number of corrections), but the key point is that this  does not induce significant variations in the performance.
Plate (c) illustrates the case where a cascading failure was simulated by turning off $50\%$ of the nodes in the network. We measure the individual entries of $\rho^P$ for a pre-determined set of nodes, which lie at a maximal distance from the sink (and are not killed). Note that the expected throughputs stabilize before the cascade, and the routes rapidly reorganize due to the failure event, when the performance regains convergent values. The entire process is perfectly decentralized, with the nodes identifying dead or non-responsive neighbors, and updating both their set of possible neighbors (Col. 2 in Table~\ref{tableNodedata}), and  self measures.

Convergence dynamics  is explicitly illustrated in Figure~\ref{figsim2}, for a dense network of $10^4$ nodes, placed on an uniform rectangular grid (uniformity merely  aids visualization). We see the gradual spreading out of the non-zero measure updates from the sink. The plates on top show the the gradient of the scalar field induced by the node measures, while those at the bottom illustrate the level sets. The voids are conglomerations of dead or non-responsive nodes. Other regions (marked  ``POOR'') comprise of nodes that are experiencing poor communication. Note that the routes tend to avoid these regions. As before, the drop probabilities are chosen randomly, and held  constant on the average  with zero mean Gaussian noise. Also, note the two color tones illustrate the possibility of simple decentralized thresholding, to autonomously segregate the network to classes which have a certain degree of connectivity to the sink, based on the convergent value of the estimated measures.

Progressive failures  are simulated in Figure~\ref{figsim3}, addressing situations  with gradual node depletions. Top row shows the failed regions in black. The network is initialized with $10^4$ nodes with energy levels distributed uniformly over a pre-specified range, leading to a realistic scenario, where nodes fail  due to to various  unmodeled effects in addition to energy spent in communication. Nodes are assumed to fail in clusters of $\sim10^2$  creating dead regions. The middle row shows packet traces  to the sink from operational nodes, and the bottom row illustrates the level sets. Note that with small number of dead regions, we can see very little ``white'' in the middle row, indicating high route utilization and low congestion. As the nodes fail, we see more  white space, indicating that most packets are now taking similar routes. Note that congestion leads to higher drop probabilities which are estimated on the fly, and  incorporated via GODDeS  in local decision-making, thus implying significant congestion-awareness.
\section{Conclusions \& Future Work}\label{sec7}
This paper introduces GODDeS: a new routing algorithm  designed to effectively exploit high quality paths in  lossy  ad-hoc wireless environments, typically with a large number of nodes.
The routing problem is modeled as an optimal control problem for a decentralized Markov Decision Process, with links  characterized by locally known packet drop probabilities that either remain constant on average or change slowly. The equivalence of this optimization problem  to that of performance maximization of an explicitly constructed  PFSA allows us to apply the theory of quantitative measures of probabilistic regular languages, and design a distributed highly efficient solution approach that attempts to minimize source-to-sink drop probabilities across the network.
 Theoretical results provide rigorous guarantees on global performance, showing that the algorithm achieves near-global optimality, in polynomial time.
It is also argued that GODDeS is significantly congestion-aware, and exploits multi-path routes optimally. Theoretical development is supported by high-fidelity network simulation.

Future work will proceed in the following directions, primarily aimed at investigating and consequently relaxing some of the key assumptions made in this paper:
\begin{enumerate}
 \item Design explicit strategies for energy and congestion awareness within the GODDeS framework. In particular, investigate the ramifications of various choices of the measure reduction factor described in Eq.~\eqref{eqfactor}.
\item Generalize the analysis to multiple sinks, which is not too difficult in view of the fact that most of the theoretical results carry over to the general case.
\item We assumed that the link-specific drop probabilities are  estimated at the nodes. Grossly incorrect estimations will translate to incorrect  routing decisions, and decentralized strategies for 
robust identification of these parameters need to be investigated at a greater depth.
\item Explicit design of implementation details such as packet headers, node data structures and pertinent neighbor-neighbor communication protocols.
\item Hardware validation  with networks of different sizes, and with induced failure situations.
\end{enumerate}

%#################################################
%#################################################
}%%endALLOWDISPLAYBREAKS  DO NOT REMOVE
%#################################################
\bibliographystyle{IEEEtran}
\bibliography{BibLib1}
\end{document}